\documentclass[11pt]{article}
\usepackage{latexsym,amssymb}
\newcommand{\ft}[2]{{\textstyle\frac{#1}{#2}}}
\def\tilde{\widetilde}

\def\1bar{1\hskip -.275cm -}
\def\2bar{2\hskip -.275cm -}
\def\3bar{3\hskip -.275cm -}

\newsavebox{\uuunit}
\sbox{\uuunit}
    {\setlength{\unitlength}{0.825em}
     \begin{picture}(0.6,0.7)
        \thinlines
        \put(0,0){\line(1,0){0.5}}
        \put(0.15,0){\line(0,1){0.7}}
        \put(0.35,0){\line(0,1){0.8}}
       \multiput(0.3,0.8)(-0.04,-0.02){10}{\rule{0.5pt}{0.5pt}}
     \end {picture}}

\makeatletter \@addtoreset{equation}{section} \makeatother

\def\bfone{\relax{\rm 1\kern-.35em 1}}

\def\bfone{\relax{\rm 1\kern-.35em 1}}
 
\begin{document}
\begin{titlepage}
\begin{flushright}
DFTT-19/2002\\
\end{flushright}
\vskip 1.5cm
\begin{center}
{\LARGE \bf  A new first order formalism for $\kappa$-supersymmetric Born Infeld actions:
  the $D3$-brane example$^\dagger$}\\ \vfill {\large
 Pietro Fr\'e and  Leonardo Modesto} \\
\vfill {
$^1$ Dipartimento di Fisica Teorica, Universit\'a di Torino, \\
$\&$ INFN -
Sezione di Torino\\
via P. Giuria 1, I-10125 Torino, Italy  }
\end{center}
\vfill
\begin{abstract}
{We introduce a new first order
formulation of  world-volume actions for
$p$-branes  with $\kappa$--supersymmetry.  In this language, which involves more auxiliary fields
compensated by more local symmetries, the action is provided by a very
compact, simple and elegant formula applicable to any supergravity
background. The $\kappa$--supersymmetry variation against which it is invariant is obtained
from the bulk supersymmetries
by means of  a projector that has a simple expression in terms of the auxiliary fields.
The distinctive feature of our formalism is that all fermion fields are hidden into the definition
of the curvatures and the action is formally the same, in terms of these  differential forms as it
would be in a purely bosonic theory. Typically our new formulation
enables one to discuss
the correct boundary actions
for non trivial  supergravity $Dp$--brane bulk solutions like the $D3$--brane solution on smooth ALE manifolds
with flux
recently constructed in the literature.}
\end{abstract}
\vspace{2mm} \vfill \hrule width 3.cm {\footnotesize $^ \dagger $
This work is supported in part by the European Union RTN contracts
HPRN-CT-2000-00122 and HPRN-CT-2000-00131.}
\end{titlepage}
\section{Introduction}
The $p$--branes and in particular $Dp$--branes are essential items in our
new non perturbative understanding of string theory after the 2nd
string revolution. There are three complementary aspects and as many
complementary approaches to the study of these solitonic excitations
of the string spectrum. Indeed they can be alternatively viewed as:
\begin{description}
  \item[a]  classical solutions of the low energy supergravity
  field equations in the bulk,
  \item[b]  world--volume gauge theories described by suitable
  world--volume actions characterized by $\kappa$--supersymmetry
  \item[c]  boundary states in the superconformal field theory
  description (SCFT) of superstring vacua.
\end{description}
The three descriptions are closely intertwined. In particular the
relation between $[a]$ and $[b]$ is the following. On one side the
world--volume actions provide the source terms that specify the
boundary conditions utilized in solving the supergravity field
equations. On the other hand, for small fluctuations of the fields
around a given classical solution, the world volume action provides
the tool to explore the gauge/gravity correspondence, namely the
pairing between certain composite operators in the world--volume
gauge theory and certain corresponding bulk modes. Alternatively when
we adopt the more abstract language of superconformal field theories,
classical string backgrounds are identified with a specific SCFT and
the brane is identified with a suitable boundary state for the same. The
world--volume action encodes the interactions within the chosen boundary
conformal field theory.
\par
From this discussion it is evident that a full command on the
world--volume actions of $p$--branes is a mostly essential weapon in
the arsenal of the modern string theorist. As mentioned above the
distinctive feature of these actions, which guides their
construction, is the so called $\kappa$--supersymmetry \cite{ksusyschwarz}.
This corresponds to the fermionic local symmetry that allows to halve the number of Fermi
fields, originally equal to the number of $\theta$--coordinates for
the relevant superspace, and obtain, on--shell, an equal number of
bosonic and fermionic degrees of freedom as required by the general
brane--scan \cite{branescan}. As it is well understood in the
literature since many years \cite{ksusytonin,Mezincescu,castdauriafre,Dall'Agata:1999wz}
the $\kappa$--supersymmetries are nothing else but suitable
\textit{chiral projections} of the original supersymmetry
transformation rules defined by supergravity. This was made
particularly evident and handy by the construction of world--volume
actions within the framework of the rheonomy approach to supergravity
\cite{castdauriafre,Dall'Agata:1999wz,0particle,allroads}. In this
geometric approach all Fermi fields are implicitly hidden in the
definition of the geometric $p$--form potentials of supergravity and
\textit{formally} the action is the same as it would be in a purely
bosonic theory. Yet it is supersymmetric and this supersymmetry, which
fixes the relative coefficients of the kinetic terms with respect to
the Wess--Zumino terms, can be shown through a simple calculation
starting from the rheonomic parametrizations of the supergravity
curvatures. In order to apply such a powerful method, the
world--volume action must be presented in first--order rather than in
second order formalism, namely {\`{a}} la Polyakov \cite{polyakov} rather than {\`{a}} la
Nambu--Goto \cite{nambugotorig}. As a consequence the rheonomic method was successfully
applied to those instances of $p$--brane actions where the Polyakov
formulation (further generalized with the introduction of an
additional auxiliary field representing the derivative of scalar
fields) did exist: in particular the string or $1$--brane \cite{castdauriafre}, the
M2--brane \cite{Dall'Agata:1999wz} and the particle or $0$--brane
\cite{0particle}. More general $Dp$--branes were so far out of reach
because of the following reason: their second order action is of the
Born--Infeld type and a suitable first order formalism for the
Born--Infeld lagrangian was not known.
\par
In this paper we just fill this gap by constructing a \textbf{new first
order formalism} that is able of generating second order actions of
the Born--Infeld type. The new formulation which, in our opinion, is
particularly compact and elegant is based on the introduction of an
additional auxiliary field, besides the world volume vielbein, and on
the enlargement of the local symmetry from the Lorentz group to the
general linear group:
\begin{equation}
  \mathrm{SO(1,d-1)}\,  \stackrel{\mbox{enlarged}}{\Longrightarrow}
  \, \mathrm{GL(d,\mathbb{R})}
\label{allargo}
\end{equation}
Within this new formalism we can easily apply the rheonomic method
and as an example we construct the $\kappa$--supersymmetric action of
a $D3$--brane. This choice is not random rather it is rooted in the main
motivations to undertake this new construction. Indeed we need a
suitable formulation of the $D3$--brane action holding true on a
generic background in order to apply it to the newly constructed
$D3$--brane bulk solutions of supergravity like that on smooth ALE
manifolds \cite{noialtrilast} or the more recent and complicated ones 
of \cite{nuovesoluzie},\cite{Billo.M}.
\par
Our paper is organized as follows. In section \ref{1storder} we
review the rheonomic formulation of $\kappa$-supersymmetry based on
an essential use of the old $1st$--order formalism. In section
\ref{thenewform} we introduce the new first order formalism and
we show how in this framework  we can recover the Born--Infeld action by eliminating
the auxiliary fields through their own equation of motion.
In section \ref{thed3examp} we apply our new machinery to the case of
the $D3$--brane and we explicitly show $\kappa$-supersymmetry.
Section \ref{outlook} contains an outlook and our conclusions.
In the appendices we have placed some important although more
technical material. Particularly relevant is appendix \ref{type2bsum} which is
a short but comprehensive summary of type IIB supergravity in the
rheonomic approach. It is entirely based on the original papers by
Castellani and Pesando \cite{castella2b,igorleo} but it contains also
some new useful results, in particular the transcription of
curvatures from the complex $\mathrm{SU(1,1)}$ basis to the real
$ \mathrm{SL(2,\mathbb{R})}$ basis and the comparison of the
supergravity field equations as written in the rheonomy approach and
as written in string text--books like Polchinsky's \cite{polchinski}.
\section{From $2nd$ to $1st$  order  and the rheonomy setup for to
$\kappa$ supersymmetry}
\label{1storder}
In this section we summarize the $1st$ order formulation of
world--volume actions and we recall their essential role in setting
up a simple, compact, rheonomic approach to $\kappa$--supersymmetry.
Then we point out the problem arising with $Dp$--branes, related to
the presence of the gauge--field $A_\mu$. In this way we establish
the need for the new first order formalism which is explained in the
next section.
\subsection{Nambu--Goto, Born--Infeld and  Polyakov kinetic actions for $p$--branes}
The $2nd$ order Nambu-Goto  action of a bosonic string
\cite{nambugotorig} that moves through a
$D$--dimensional space--time endowed with a metric $g_{\underline{\mu \nu }}$, is simply given by the area of the
world--sheet swept by the string. Namely we have:
\begin{equation}
  \mathcal{A}^{Nambu \,\, Goto}_{string} = \int \, d^2\xi \,
  \sqrt{-\mbox{det}\,
  G_{\mu\nu} }
\label{nambugstrig}
\end{equation}
where:
\begin{equation}
  G_{\mu\nu} \equiv \partial _\mu X^{\underline{\mu}} \, \partial _\nu
  X^{\underline{\nu}}\, g_{\underline{\mu \nu} }
\label{pulbaccus}
\end{equation}
denotes the pull-back of the bulk metric $g_{\underline{\mu \nu}} (X)$ onto
the world--sheet. Such an action admits a
straightforward generalization to the case of a $p$-brane, the area
of the world--sheet being replaced by the value of the
$d=p+1$--dimensional
world--volume:
\begin{equation}
  \mathcal{A}^{Nambu \,\, Goto}_{p-brane} = \int \, d^d\xi \,
  \sqrt{-\mbox{det}\,
  G_{\mu\nu} }
\label{nambugbran}
\end{equation}
As it is well known from the literature and thoroughly discussed in
recent string theory textbooks \cite{polchinski}, the kinetic part of
$Dp$--brane  actions is provided  by a further generalization of the
Nambu--Goto action (\ref{nambugbran}) where the symmetric matrix $G_{{\mu \nu}
}$ is modified by the addition of an antisymmetric part $ F_{\mu
\nu}$ that represents the field strength of a world volume gauge field $A_\mu$:
\begin{eqnarray}
\mbox{for $D$--branes} & G_{\mu \nu}
 \mapsto & G_{\mu \nu} + F_{\mu \nu} \nonumber\\
\mbox{where} & F_{\mu \nu}= & \ft{1}{2} \left( \partial_\mu A_\nu  - \partial _\nu  A_\mu \right)
\label{crucca}
\end{eqnarray}
Seen from a different perspective the resulting second order action:
\begin{equation}
  \mathcal{A}^{kinetic}_{D-brane} = \int \, d^d\xi \, \sqrt{
  -\mbox{det} \left( G_{\mu \nu } + F_{\mu \nu } \right) }
\label{borninfeld}
\end{equation}
is a generalization of the Born-Infeld \cite{borninfeld} action of non linear
electromagnetism. Indeed the latter was early shown to be the effective action for
the zero mode gauge field of an open string theory
\cite{tseytlinaction}.
\par
In the context of superstrings and in the analysis of
$Dp$--brane systems the important issue is to write world--volume
actions that possess both reparametrization invariance and $\kappa$-supersymmetry
\cite{ksusyschwarz}. The former is needed to
remove the unphysical degrees of freedom of the bosonic sector, while the
latter removes the unphysical fermions. In this way we end up with an equal number of
physical bosons and physical fermions as it is required by supersymmetry.
As widely discussed in the literature \cite{ksusytonin,castdauriafre,Dall'Agata:1999wz,allroads}
the appropriate $\kappa$-supersymmetry transformation rules are
nothing else but the supersymmetry transformation rules of the bulk
supergravity background fields with a special supersymmetry parameter
$\epsilon$ that is projected onto the brane. For those $\kappa$-supersymmetric branes where
the gauge field strength $F_{\mu \nu }$ is not required (for example the string itself or the M2--brane)
such a projection is realized by imposing that the spinor $\epsilon$
satisfies the following condition:
\begin{equation}
  \epsilon = \ft {1}{2} \left( 1 + \left( {\rm i}\right) ^{d+1}\, \ft{1}{d!} \, \Gamma_{\underline{a_1\dots a_d}}
  V^{\underline{a_1}}_{i_1} \, \dots \,
  V^{\underline{a_d}}_{i_d}\,\epsilon^{i_1\dots i_d} \right)
  \epsilon
\label{projparam}
\end{equation}
where $\Gamma_{\underline{a}}$ are the gamma matrices in
$D$--dimensions and $V^{\underline{a}}_{m}$ are the component of
the bulk vielbein $V^{\underline{a}}$ onto a basis of world-volume
vielbein $e^m$. Explicitly we write
\begin{equation}
  V^{\underline{b}}_m \, e^m = \varphi^* \, \left[ V^{\underline{b}} \right]
\label{pullobacco}
\end{equation}
where $\varphi^* \, \left[ V^{\underline{b}} \right]$ denotes  the pull--back of the bulk vielbein on the world volume,
\begin{equation}
  \varphi \, : \, \mathcal{W}_d \, \hookrightarrow \,  \mathcal{M}_D
\label{inietto}
\end{equation}
being the injection map of the latter into the former.
For all other branes with a full--fledged Born Infeld type of action
the projection (\ref{projparam}) becomes more complicated and
involves also $F_{\mu \nu }$.
\par
Certainly one can address the problem of $\kappa$--supersymmetrizing
the $2nd$--order action (\ref{borninfeld}) and this programme was
carried through in the literature to some extent
\cite{gotebordbrane,tonindbrane}. Yet due to the highly
non linear structure of such a bosonic action its
supersymmetrization turns out to be quite involved. Furthermore  the
geometric structure is not transparent  and
any modification is very difficult and obscure in such an approach.
\par
On the contrary it was shown in \cite{castdauriafre} and recently
illustrated with the case of the M2--brane in
\cite{Dall'Agata:1999wz} and with the case of the $0$--brane in
four--dimensions in \cite{0particle} that by using a first order
formalism on the world volume  the implementation of
$\kappa$--supersymmetry is reduced to an almost trivial matter once the rheonomic
parametrizations, consistent with superspace
Bianchi identities,  are given for all the the curvatures of the bulk background fields.
It follows that an appropriate first order formulation of the
Born--Infeld  action (\ref{borninfeld}) is an essential step for
an easy and elegant approach to $\kappa$--supersymmetric $Dp$--brane
world volume actions that are also sufficiently versatile to adapt to all type  of
bulk backgrounds.
\par
The first order formulation of the Nambu--Goto action
(\ref{nambugbran}) is the Polyakov action for $p$--branes:
\begin{equation}
  \mathcal{L}^{Polyakov}_{p-brane} = \frac {1}{2(d-1)} \, \int \, d^d\xi
  \, \sqrt{ - \, \mbox{det}\, h_{\mu \nu }} \, \left\{ \, h^{\rho \sigma
  } \, \partial _\rho  X^{\underline{\mu} } \, \partial _\sigma  X^{\underline{\nu}
  } \, g_{\underline{\mu \nu} } + \left( d-2 \right) \right\}
\label{Polyak}
\end{equation}
where the auxiliary field $h_{\rho \sigma }$ denotes the world--volume
metric. Varying the action (\ref{Polyak}) with respect to $\delta h_{\rho \sigma
}$ we obtain the equation:
\begin{equation}
  h_{\rho \sigma } = G_{\rho \sigma }
\label{identh}
\end{equation}
and substituting (\ref{identh}) back into (\ref{Polyak}) we retrieve
the second order action (\ref{nambugbran}).
\par
The Polyakov action (\ref{Polyak}) is not yet in a suitable form for
a simple rheonomic implementation of $\kappa$--supersymmetry but can
be easily converted to such a form. The required steps are:
\begin{enumerate}
  \item replacing the world--volume metric $h_{\mu \nu }(\xi)$ with a
  world--volume vielbein $e^i=e^i_\rho \, d\xi^\rho$,
  \item using a first order formalism also for the derivatives of
  target space coordinates $X^{\underline{\mu}}$ with respect to the
  world volume coordinates $\xi^\rho$,
  \item write everything only in terms of flat components
  both on the world volume and in the target space.
\end{enumerate}
This programme is achieved by introducing an auxiliary $0$--form
field $\Pi^{\underline{a}}_i(\xi)$ with an index $\underline{a}$
running in the vector representation of $\mathrm{SO(1,D-1)}$ and a
second index $i$ running in the vector representation of
$\mathrm{SO(1,d-1)}$ and writing the action \footnote{The need of a cosmological term for p-brane actions with $p \neq 1$ was first noted by Tucker and Howe in \cite{Howe1}.} :

\begin{eqnarray}
  \mathcal{A}^{kin}[d] & = &  \int_{\mathcal{W}_d} \, \left [ \, \Pi^{\underline{a}}_j
  \,
  V^{\underline{b}}\, \eta_{\underline{ab}} \, \wedge  \eta^{ji_1} \, e^{i_2} \, \wedge \, \dots
  \, \wedge \, e^{i_d} \, \epsilon_{i_1 \dots i_d} \right.\nonumber\\
  &&\left. - \frac {1}{2d} \,\left (   \, \Pi^{\underline{a}}_i \,
  \Pi^{\underline{b}}_j \, \eta^{ij} \, \eta_{\underline{ab}} \, + \,
  d-2 \, \right)  \, e^{i_1} \, \wedge \, \dots \, \wedge \,
  e^{i_d} \, \epsilon_{i_1 \dots i_d} \right ]
\label{kin}
\end{eqnarray}
The variation of (\ref{kin}) with respect to $\delta
\Pi^{\underline{a}}_j$ yields an equation that admits the unique
algebraic solution:
\begin{equation}
  V^{\underline{a}} \vert_{\mathcal{W}_d} = \Pi^{\underline{a}}_i \,
  e^i
\label{urbano}
\end{equation}
Hence the $0$-form $\Pi^{a}_i$ is identified with the
intrinsic components along the world--volume vielbein $e^i$ of the
the bulk vielbein $V^{\underline{a}}$ pulled-back onto the world
volume. In other words the field $\Pi^{a}_i$ is identified by its own
field equation with the field $V^{\underline{a}}_i$ defined in eq.
(\ref{pullobacco}).
On the other hand with the chosen numerical coefficients the
variation of (\ref{kin}) with respect to the world--volume vielbein
$\delta e^i$ yields another equation with the unique algebraic
solution:
\begin{equation}
  \Pi^{a}_i \, \Pi^{b}_j \, \eta_{\underline{ab}} = \eta_{ij}
\label{flattopullo}
\end{equation}

which is the flat index transcription of eq.(\ref{identh})
identifying the world--volume metric with the pull-back of the bulk
metric. Hence eliminating all the auxiliary fields via their own
equation of motion the first order action (\ref{kin}) becomes
proportional to the $2nd$ order Nambu--Goto action
(\ref{nambugbran}). The first order form (\ref{kin}) of the kinetic action is the best
suited one to discuss $\kappa$-supersymmetry. To illustrate this
point we briefly consider the case of the $\mathrm{M2}$--brane
\subsection{$\kappa$-supersymmetry and the example of the M2--brane}
In the case of the M2--brane in eleven dimensions the world--volume
is three--dimensional and the complete action is simply given by the
kinetic action (\ref{kin}) with $d=3$ plus the Wess-Zumino term,
namely the integral of the $3$--form gauge field $A^{[3]}$.
Explicitly we have
$\mathrm{W}_3$:
\begin{equation}
  \mathcal{A}_{M2} = \mathcal{A}^{kin}[d=3] \, - \, {\bf q} \,\int_{\mathcal{W}_3} \, A^{[3]}
\label{M2action}
\end{equation}
where ${\bf q}=\pm 1$ is the charge of the M2--brane. As explained in
\cite{Dall'Agata:1999wz}, the background fields, namely the bulk elfbein
$V^{\underline{a}}$ an the bulk three--form $A^{[3]}$ are superspace
differential forms which are assumed to satisfy the Bianchi consistent
rheonomic parametrizations of $D=11$ supergravity as originally given in
\cite{riccapiet,pietd11}. Hence, although implicitly, the
action functional (\ref{M2action}) depends both on $11$ bosonic
fields, namely the $X^{\underline{\mu}}(\xi)$ coordinates of bulk
space--time, and on $32$ fermionic fields
$\theta^{\underline{\alpha}}(\xi)$, forming an $11$--dimensional Majorana
spinor. A supersymmetry variation of the background fields is
determined by the rheonomic parametrization of the curvatures and has
the following explicit form:
\begin{eqnarray}
\delta \, V^{\underline{a}} &=& \mbox{\rm i} {\bar \epsilon} \, \Gamma^{\underline{a}} \,
\Psi, \nonumber\\
\label{ordsusy}
\delta \,\Psi &=& {\cal D} { \epsilon} \, -
 \frac{\rm i}{3} \left( \Gamma^{\underline{b}_1\underline{b}_2\underline{b}_3} \,
F_{\underline{a}\underline{b}_1\underline{b}_2\underline{b}_3} - \frac{1}{8}
\Gamma_{\underline{a}\underline{b}_1\dots\underline{b}_4} \,
F^{\underline{b}_1\dots\underline{b}_4} \, \right) \, \epsilon\, V^{\underline{a}}, \\
\delta \, A^{[3]} & = & - \mbox{\rm i} {\bar \epsilon} \, \Gamma^{\underline{a}\underline{b}}
\Psi \, \wedge \, V_{\underline{a}} \, \wedge \, V_{\underline{b}}
\label{d11susies}
\end{eqnarray}
where $\Psi$ is the gravitino $1$--form,
$F_{\underline{a_1,\dots,a_4}}$ are the intrinsic components of the
$A^{[3]}$ curvature and $\epsilon$ is a $32$--component spinor
parameter. Essentially a supersymmetry transformation is a translation
of the fermionic coordinates $\theta \mapsto \theta + \epsilon$,
namely at lowest order in $\theta$ it is just such a translation.
With such an information the $\kappa$--supersymmetry invariance of the
action (\ref{M2action}) can be established through a two--line
computation, using the so called $1.5$--order formalism. Technically
this consists of the following. In the action (\ref{M2action}) we vary only the background
fields $V^{a},A^{[3]}$ with respect to the supersymmetry
transformations (\ref{d11susies}) and, after variation, we use the
first order field equations (\ref{urbano}), (\ref{flattopullo}). The
action is supersymmetric if all the generated terms, proportional to
the gravitino $1$--form $\Psi$ cancel against each other. This
does not happen for a generic $32$--component spinor $\epsilon$ but
it does if the latter is of the form:
\begin{eqnarray}
\epsilon &=& \frac{1}{2} \left( 1 -{\bf q}\mbox{\rm i} {\bar \Gamma}
\right)\, \kappa, \nonumber \\
{\bar \Gamma} &\equiv & \frac{\epsilon^{ijk}}{3! } \Gamma _{ijk} =
\frac{\epsilon^{ijk}}{3! } \Pi_i{}^{\underline{a}}  \Pi_j{}^{\underline{b}}
\Pi_k{}^{\underline{c}} \Gamma_{\underline{a}\underline{b}\underline{c}},
\label{projector}
\end{eqnarray}
where $\kappa$ is another spinor. Eq.(\ref{projector}) corresponds to
the anticipated projection (\ref{projparam}) which halves the spinor
components. It follows that of the $32$ fermionic degrees of freedom
$16$ can be gauged away by $\kappa$--supersymmetry. The remaining
$16$ are further reduced to $8$ by their field equation which is
implicitly determined by the action (\ref{M2action}). As one sees,
once the M2--action is cast into the first order form
(\ref{M2action}),
$\kappa$-supersymmetry invariance can be implemented in an extremely
simple and elegant way that requires only a couple of algebraic
manipulations with gamma matrices.
\par
The example of the M2--brane is generalized to all other instances of
$p$--branes where the world volume spectrum includes just the scalars
(=target space coordinates) and their fermionic partners.
\subsection{With $Dp$--branes we have a problem: the world--volume gauge field
$A^{[1]}$ }
It is clear from what we explained above that to deal with
$\kappa$-supersymmetry in an easy way we need a first order formulation of the
action. Yet in the case of $Dp$--branes there is a new problem
intrinsically related to the new structure of the Born--Infeld action
(\ref{borninfeld}) which, differently from the pure Nambu--Goto action
(\ref{nambugbran}) cannot be recast into a first order form of
type (\ref{kin}) \footnote{Actually a partial first order formalism was already introduced in the literature for Dp-branes \cite{Dima1} in the context of the superembedding approach initiated by Kharkov group and extensively developed also in collaborations with the Padua group and ather groups \cite{Howe}.
In particular in \cite{Dima2} an action with a partial first order formalism was introduced in the sense that there is an auxiliary $F_{ij}$ field for the gauge degres of freedom but the action is ``second order'' in the brane coordinates $x$ and $\theta$, which enter through the pullback of the target space supervielbine $E^{a}$.}.
The solution of this problem is found through a
procedure which is very frequent and traditional in Physics. Indeed,
when a certain formulation of a theory cannot be generalized to a
a wider scenario including  additional fields it
usually means that there is a second formulation of the same theory
which is equivalent to the former in the absence of the new fields, but
which, differently from the former, can incorporate them in a natural
way. Typical example is the relation of Cartan's formulation of General Relativity
in terms of vielbein and spin connection with respect to the standard metric formulation. Although
they are fully
equivalent in the absence of fermions, yet the former allows the coupling to spinors while the latter
does not.
The present case is similar. It turns out that there is a new, so far unknown, first
order formulation of world-volume actions which, in the absence of world--volume
gauge fields is fully equivalent to the formulation of eq. (\ref{kin}). Yet
world--volume $1$--forms can be naturally included in the new
formalism while they have no place in the old. In full analogy with
other examples of the same logical process the new formalism  relies
on the addition of a new auxiliary field and a new symmetry. The new
field is a $0$--form rank $2$ tensor $h_{ij}$ that is identified with
the intrinsic components of the pulled-back bulk metric along a
reference world--volume vielbein $e^i$. The new symmetry is the
independence of the action from the choice of the reference vielbein.
Explicitly this means the following. Let $K^i_{\phantom{i}j}(x)$ be a
generic $d \times d$ matrix depending on the world--volume point. The
new action we shall construct will be invariant against the
local transformation:
\begin{eqnarray}
e^i & \mapsto & K^i_{\phantom{i}j} \, e^j \nonumber\\
h^{ij} & \mapsto &  \left( K^{-1}\right)^i_{\phantom{i}i^\prime}\,
\left( K^{-1}\right) ^j_{\phantom{i}j^\prime}\, h^{i^\prime j^\prime} \, \left(
\mbox{det} K \right)
\label{transfa}
\end{eqnarray}
accompanied by suitable transformation of the other fields. The
above symmetry generalizes the local Lorentz invariance of the
previously known first order $p$--brane actions. Indeed, being
generic, the matrix $K$ can in particular be an element of the
Lorentz group $K\in \mathrm{SO(1,d-1)}$. In this case there is no
novelty. However $K$ can also be a representative of a non
trivial equivalence class of the coset $\mathrm{GL(d,\mathbb{R})}/ \mathrm{SO(1,d-1)}$. This
latter is precisely
parametrized by arbitrary symmetric matrices. Hence the additional
degrees of freedom introduced by the new auxiliary field $h^{ij}$ are
taken away by the enlargement of the local symmetry from
$\mathrm{SO(1,d-1)}$ to $\mathrm{GL(d,\mathbb{R})}$.
\section{The new first order formalism}
\label{thenewform}
In the next subsection \ref{newforma} we describe the new formalism as an alternative
to the action (\ref{kin}). Then in subsection \ref{includo} we show how
it allows the inclusion of world volume gauge fields and provides a
first order formulation of the Born--infeld action (\ref{borninfeld}).
\subsection{An alternative to the Polyakov action for  $p$--branes}
\label{newforma}
To begin with we consider a world--volume Lagrangian of the following
form:
\begin{eqnarray}
\mathcal{L} & = & \Pi^{\underline{a}} _i \, V^{\underline{b}} \, \eta_{\underline{ab}}
 \, \eta^{i\ell_1} \, \wedge e^{\ell_2} \, \wedge \, \dots \wedge e^{\ell_d} \, \epsilon_{\ell_1 \dots \ell_d}
  + a_1 \, \Pi^{\underline{a}} _i \, \Pi^{\underline{b}}
_j \, \eta_{\underline{ab}} \, h^{ij} \, e^{\ell_1} \, \wedge \, \dots
\wedge e^{\ell_d} \, \epsilon_{\ell_1 \dots \ell_d} \nonumber\\
&&+ a_2 \, \left( \mbox{det}\, h\right) ^{-\alpha} \,  e^{\ell_1} \, \wedge \, \dots
\wedge e^{\ell_d} \, \epsilon_{\ell_1 \dots \ell_d}
\label{primesemp}
\end{eqnarray}
where $a_1 \, , \, a_2 \, , \, \alpha $ are real parameters to be
determined and the other notations are recalled in eq.(\ref{definizie}) of
appendix \ref{notazie}.
\par
Performing the $\delta \, \Pi^{\underline{a}}_i$ variation of the
Lagrangian (\ref{primesemp}) we obtain:
\begin{equation}
  \eta_{\underline{ab}} \, V^{\underline{b}}_m \, \eta^{i\ell_1} \,
  \epsilon^{m\ell_2 \dots \ell_d} \, \epsilon_{\ell_1 \dots \ell_d}\,
  +\, 2 \, (d!) \, a_1 \, \eta_{\underline{ab}} \,
  \Pi^{\underline{b}}_j \, h^{ij} = 0
\label{Piequa}
\end{equation}
\par\noindent
If we choose:
\begin{equation}
  a_1 = - \frac{1}{2 \, d}
\label{a1fix}
\end{equation}
then equation (\ref{Piequa}) is solved  by:
\begin{equation}
  \Pi^{\underline{b}}_m = V^{\underline{b}}_i \, \eta^{ip} \, \left( h^{-1} \right)
  _{pm}
\label{Pisolvo}
\end{equation}
Let us then introduce the following three $d \, \times \, d$
matrices:
\begin{equation}
  \gamma_{ij} = \Pi^{a}_i \, \Pi^{b}_j \, \eta_{\underline{ab}} \quad
  ; \quad G_{ij} = V^{\underline{a}}_i \, V^{\underline{b}}_j \, \eta_{\underline{ab}} \quad
  ; \quad \widehat{G} = \eta \, G \, \eta
\label{trematre}
\end{equation}
The solution (\ref{Pisolvo}) of the field equation (\ref{Piequa})
implies that:
\begin{equation}
  \gamma = \left( h^{-1}\right) ^{T} \, \eta \, G \, \eta \,  h^{-1} = \left( h^{-1}\right) ^{T} \, \widehat{G} \,
  h^{-1}
\label{gammasolvo}
\end{equation}
Next let us consider the variation of the action (\ref{primesemp})
with respect to the symmetric matrix $h^{ij}$. In matrix form such a
variational equation reads as follows:
\begin{equation}
  a_1 \, \gamma \, - \, a_2 \, \alpha \, h^{-1} \, \left( \mbox{det} \, h \right)
  ^{-\alpha} = 0
\label{hmatrequa}
\end{equation}
Setting:
\begin{equation}
  a_2 =  \frac{a_1}{\alpha} = -\frac{1 }{2 \, d \, \alpha}
\label{a2fix}
\end{equation}
eq.(\ref{hmatrequa}) reduces to
\begin{equation}
  \gamma= h^{-1} \, \left(\mbox{det}\, h \right) ^{-\alpha}
\label{gammanewequa}
\end{equation}
which can be solved by the ansatz:
\begin{equation}
  h= \gamma^{-1} \, \left( \mbox{det} \, \gamma \right) ^{\beta}
\label{betansaz}
\end{equation}
provided:
\begin{equation}
  \beta = \frac{\alpha}{d \, \alpha + 1}
\label{betavalue}
\end{equation}
On the other hand from eq.(\ref{gammasolvo}) we get:
\begin{equation}
  \mbox{det}\, \gamma \, = \, \mbox{det}G \, \left( \mbox{det} \, h\right)
  ^{-2}
\label{intermed}
\end{equation}
so that:
\begin{equation}
  h= h \, \widehat{G}^{-1} \, h \left( \mbox{det} G\right) ^\beta \, \left(
  \mbox{det}h \right) ^{-2\beta}
\label{newhequaz}
\end{equation}
Eq.(\ref{newhequaz}) can be solved by the ansatz:
\begin{equation}
  h = \widehat{G} \, \left( \mbox{det} \, G\right) ^p
\label{newsolvo}
\end{equation}
provided:
\begin{equation}
  p= -\frac{\alpha}{d \, \alpha -1}
\label{pfix}
\end{equation}
Combining the last two results we have the final solution for the two auxiliary fields $h$ and $\gamma$:
\begin{equation}
  h= \widehat{G} \, \left( \mbox{det} \, G \right) ^p \quad ; \quad \gamma =
  \widehat{G}^{-1} \, \left( \mbox{det} G \right) ^{-2p}
\label{Gsolvo}
\end{equation}
in terms of $G$ which is just the pull-back of the bulk metric onto the
world volume, expressed in flat components with respect to an arbitrary reference vielbein
$e^\ell$ that lives on $ \mathcal{W}$.
\par
Using eq.(\ref{Gsolvo}) we can rewrite the action (\ref{primesemp})
in second order formalism. The basic observation is that after
implementation of the first order field equations the three terms
appearing in (\ref{primesemp}) become all proportional to the same
term, namely $\left( \mbox{det}\, G\right) ^{-p} \, \mbox{det e} \, d^d\xi$, having named
$\xi$ the world volume coordinates. Indeed we have:
\begin{eqnarray}
\left( \mbox{ det} h\right) ^{-\alpha} \, e^{\ell_1} \, \wedge \, \dots \,
e^{\ell_d} \, \epsilon_{\ell_1 \dots \ell_d}
& = & d! \, \left( \mbox{det}\, G\right) ^{-p} \, \mbox{det e} \, d^d\xi \nonumber\\
\eta_{\underline{ab}}\, \Pi^{\underline{a}}_i \, \Pi^{\underline{b}}_{j} \, h^{ij} \, e^{\ell_1} \, \wedge \, \dots \,
e^{\ell_d} \, \epsilon_{\ell_1 \dots \ell_d} & = & d \, d! \,
\left( \mbox{det}\, G\right) ^{-p} \, \mbox{det e} \, d^d\xi \nonumber\\
\Pi^{\underline{a}}_i \, V^{\underline{b}}\, \eta_{\underline{ab}}
 \, \eta^{i\ell_1} \, \wedge e^{\ell_2} \, \wedge \, \dots \wedge e^{\ell_d} \, \epsilon_{\ell_1 \dots
 \ell_d} & = & d! \, \left( \mbox{det}\, G\right) ^{-p} \, \mbox{det e} \, d^d\xi
\label{tuttiugual}
\end{eqnarray}
Hence the Lagrangian (\ref{primesemp}) becomes:
\begin{eqnarray}
  \mathcal{L} &=& (d-1)! \,  \left( \mbox{det} G \,
  \right)^{-p}\, \mbox{det e} \, d^d\xi \nonumber\\
  & = &  (d-1)! \, \frac{1}{2p} \left( \mbox{det} G_{\mu\nu} \,
  \right)^{-p}\, \left( \mbox{det e}\right) ^{2p+1} \, d^d\xi
\label{metamorfo}
\end{eqnarray}
the second identity following from:
\begin{eqnarray}
  G_{ij} &=& V^{\underline{a}}_{\underline{\mu}} \,
  V^{\underline{b}}_{\underline{\nu}} \, \eta_{\underline{ab}} \,
  \partial_\mu X^{\underline{\mu}} \, \partial_\nu
  X^{\underline{\nu}}\, e^\mu_i \, e^\nu_j \, \, = \, \,
  \underbrace{g_{\underline{\mu\nu}} \, \partial_\mu X^{\underline{\mu}} \, \partial_\nu
  X^{\underline{\nu}}}_{G_{\mu\nu} } \,  e^\mu_i \, e^\nu_j
  \nonumber\\
  & \Downarrow & \nonumber\\
  \mbox{det}\, G_{ij} & = & \left( \mbox{det} \, G_{\mu \nu } \,
  \right)\, \left( \mbox{det}\, e \right) ^{-2}
\label{moltecose}
\end{eqnarray}
where $G_{\mu \nu }$ denotes the pull--back of the bulk space--time
metric $g_{\underline{\mu \nu} }$ onto the world--volume of the
brane.
\par
If we choose:
\begin{equation}
  p= -\frac{1}{2}\quad \Rightarrow \quad \alpha=  \frac{1}{d-2}
\label{sceltadip}
\end{equation}
then the original world-volume lagrangian (\ref{primesemp}), already
transformed to the second order form (\ref{metamorfo}) becomes proportional to the Nambu--Goto lagrangian :
\begin{equation}
  \mathcal{L}= (d-1)! \,  \sqrt{ \mbox{det}\, G_{\mu \nu
  } } \, d^d \xi
\label{nambugoto}
\end{equation}
In this way the reference vielbein $e^i_\mu$ has disappeared from the
lagrangian. This result is supported by the calculation of the
variation in $\delta e^k$ of the first order action
(\ref{primesemp}). After variation and substitution of the result for
the first order equations $\delta \Pi^{\underline{a}}_i$ and $\delta
h_{ij}$ all terms are already Kronecker deltas proportional to
$\mbox{det}\, G$. With the choice $p=-1/2$ all terms in this stress
energy tensor cancel identically.
\par
Note also that if the transformation (\ref{transfa}) is completed by
setting:
\begin{equation}
  \Pi^{\underline{a}}_i \mapsto \, K^i_{\phantom{j}k} \, \eta^{k\ell}
  \, \Pi^{\underline{a}}_\ell \, \left( \mbox{det} K\right) ^{-1}
\label{Picaptrans}
\end{equation}
it becomes an exact local symmetry of the action (\ref{primesemp}).
\par
In this way we have shown how the standard first order formalism for
the Nambu--Goto action can be replaced by a new first order formalism
involving the additional field $h_{ij}$. So far the matrix $h$ was
chosen to be symmetric. Including world--sheet vector fields
corresponds to the generalization of the above construction to the
case where $h$ has also an antisymmetric part.
\subsection{Inclusion of a world--volume gauge field and the Born
Infeld action in first order formalism}
\label{includo}
We consider a modification of the first order action
(\ref{primesemp}) of the following form
\begin{eqnarray}
\mathcal{L} & = & \Pi^{\underline{a}} _i \, V^{\underline{b}} \, \eta_{\underline{ab}}
 \, \eta^{i\ell_1} \, \wedge e^{\ell_2} \, \wedge \, \dots \wedge e^{\ell_d} \, \epsilon_{\ell_1 \dots \ell_d}
  + a_1 \, \Pi^{\underline{a}} _i \, \Pi^{\underline{b}}
_j \, \eta_{\underline{ab}} \, h^{ij} \, e^{\ell_1} \, \wedge \, \dots
\wedge e^{\ell_d} \, \epsilon_{\ell_1 \dots \ell_d} \nonumber\\
&&+ a_2 \, \left [ \mbox{det}\,\left(  h^{-1} + \mu \mathcal{F}\right) \right] ^\alpha \,  e^{\ell_1} \, \wedge \, \dots
\wedge e^{\ell_d} \, \epsilon_{\ell_1 \dots \ell_d}\nonumber\\
&& + a_3 \mathcal{F}^{ij} \, F^{[2]} \, \wedge \, e^{\ell_3} \, \wedge \,
\dots \, \wedge \,e^{\ell_d} \, \epsilon_{ij\ell_3 \dots \ell_d} + \underbrace{WZT}_{\mbox{Wess Zumino terms}}
\label{duesemp}
\end{eqnarray}
where
\begin{equation}
  F^{[2]} \equiv dA^{[1]}
\label{FdefidA}
\end{equation}
is the field strength of a world--volume $1$--form gauge field,
$\mathcal{F}_{ij} = - \mathcal{F}_{ji}$ is an antisymmetric $0$--form auxiliary field  and
$a_3$ is a further numerical coefficient to be determined. Furthermore
$WZT$ denotes the Wess--Zumino terms, i.e. the integrals on the world volume of various combinations of the
Ramond--Ramond $p$--forms. These terms depend on the type of $Dp$--brane considered and will be discussed later in
the case of the $D3$--brane.
\par
Performing the $\delta \Pi^{\underline{a}}_i$ variation we obtain :
\begin{equation}
(d-1)! \, [\eta_{\underline{ab}} \, V^{\underline{a}}_l \, \eta^{il}+2a_1 d \, \eta_{\underline{ab}} \, \Pi^{\underline{a}}_j \, h^{ij}] \, = \, 0
\end{equation}
that is solved by :
\begin{equation}
\Pi^{\underline{a}}_j \, = \, -\frac{1}{2 \, d \,a_1} \, V^{\underline{a}}_m \,(h^{-1})^{m} \, _j
\label{deltaPi}
\end{equation}
and :
\begin{equation}
  \gamma = \, \frac{1}{(2 \, d \,a_1)^{2}} \, h^{-1} \, \widehat{G} \, h^{-1}
\label{asbefor}
\end{equation}
Varying in $\delta h_{ij}$ we also obtain a result similar to what we had
before, namely:
\begin{equation}
  a_1 \, \gamma \, - \, a_2 \alpha \, h^{-1} \, \left( h^{-1} + \mu  \mathcal{F}\right)
  ^{-1}_S \, h^{-1} \, \left[ \mbox{det} \, \left( h^{-1} +
 \mu  \mathcal{F}\right) \right]^\alpha = 0
\label{similarbefor}
\end{equation}
where the suffix $S$ denotes the symmetric part of the matrix to
which it is applied.\par
From the variotion in $\delta \mathcal{F}_{ij}$ we obtain instead:
\begin{equation}
  - d! \, a_2 \, \alpha \, \mu \, \left(  h^{-1} + \mu \mathcal{F}\right) ^{-1}_A \, \left[ \mbox{det}\, \left (h^{-1} +\mu \mathcal{F}\right) \right]^\alpha + 2 \, (d-2)! \, a_3 F = 0
\label{Fvariation}
\end{equation}
where the suffix $A$ denotes the antisymmetric part of the matrix to
which it is applied and where $F$ is the antisymmetric matrix
$F_{ij}$ of flat components of the field strength $2$--form:
\begin{equation}
  F^{[2]} = F_{ij} \, e^i \, \wedge \, e^j
\label{Famat}
\end{equation}
Hence from $\delta h_{ij}$ and $\delta \mathcal{F}_{ij}$ we get:
\begin{eqnarray}
\frac{2(d-2)! \, a_3 \, }{d! \, a_2 \, \alpha \, \mu}   F & = & \left( h^{-1} + \mu \mathcal{F}\right) ^{-1}_A \, \left[ \mbox{det} \, \left( h^{-1} +
 \mu \mathcal{F} \right) \right] ^{\alpha }
\nonumber\\
\frac{1}{4 \, d^{2} \, a_1 \, a_2 \, \alpha} \widehat{G}  & = & \left( h^{-1} +  \mu \mathcal{F}\right) ^{-1}_S \, \left[ \mbox{det} \, \left( h^{-1} +
 \mu \mathcal{F} \right) \right] ^{\alpha }
\label{2eque}
\end{eqnarray}
Summing the two eq.s (\ref{2eque}) together we obtain:
\begin{equation}
\frac{2(d-2)! \, a_3 \, }{d! \, a_2 \, \alpha \, \mu} \, F+\frac{1}{4 \, d^{2} \, a_1 \, a_2 \, \alpha} \, \widehat{G} = \left( h^{-1} + \mu \mathcal{F}\right) ^{-1} \, \left[ \mbox{det} \, \left( h^{-1} + \mu \mathcal{F} \right) \right] ^{\alpha }
\label{unequa}
\end{equation}
which can be uniquely solved by:
\begin{equation}
  h^{-1} + \mu \mathcal{F} = \left(a \, \widehat{G} + b \, F\right) ^{-1} \, \left[ \mbox{det}
  \left(a \, \widehat{G} + b \, F \right) \right] ^\beta \quad ; \quad \beta = \frac{\alpha }{\alpha  \, d - 1}
\label{h+fstort}
\end{equation}
where :
\begin{equation}
a=\frac{1}{4 \, d^{2} \, a_1 \, a_2 \, \alpha} \quad ; \quad
b=\frac{2(d-2)! \, a_3 \, }{d! \, a_2 \, \alpha \, \mu}
\label{def a,b}
\end{equation}
The coefficients $a_1$, $a_2$, $a_3$ are redundant since they can be reabsorbed
into the definition of $\Pi^{\underline{a}}_{j}$, $h$ and $\mathcal{F}$; so we fix them by imposing :
\begin{equation}
a_1 = -\frac{1}{2d} \quad ; \quad
a = 1 \quad ; \quad
b = -\frac{1}{\mu}
\label{def a1,a2,a3}
\end{equation}
Hence using (\ref{def a,b}) and (\ref{def a1,a2,a3}) we obtain :
\begin{equation}
a_2 = -\frac{1}{2d \alpha} \quad ; \quad
a_3 = -\frac{d! a_2 \alpha}{2 (d-2)!} = \frac{d-1}{4}
\label{valori a2,a3}
\end{equation}
At this point everything proceeds just as in the previous case. Indeed inserting eq.s (\ref{asbefor}), (\ref{deltaPi})
 back into the action (\ref{duesemp}) we obtain :
\begin{eqnarray}
[ \left( -\frac{(d-1)!}{2 \, d \, a_1} \, + a_1 \, d! \right) \frac{1}
{(2 \, d \, a_1)^2} \, Tr( h^{-1} \, \widehat{G})  \, +2 \,  a_3 \, (d-2)! \, \mathcal{F}^{ij} \, {F}_{ij}    \, + \nonumber     \\
+a_2 \, d! \, [\mbox{det} \, \left( h^{-1}
+ \mu \mathcal{F} \right)]^{\alpha} ] \,
\mbox{det} \, e \, d^d\xi
\label{eqBI}
\end{eqnarray}
Using (\ref{def a1,a2,a3}) and(\ref{valori a2,a3}) eq. (\ref{eqBI}) becomes :
\begin{equation}
\{ \frac{(d-1)!}{2} [Tr(h^{-1} \widehat{G}) \, - \, Tr(\mathcal{F}F)] - \frac{(d-1)!}{2\alpha}[\mbox{det} \, \left( h^{-1} + \mu \mathcal{F} \right)]^{\alpha}  \}\mbox{det} \, e \, d^d\xi
\label{eqBI 1}
\end{equation}
Now we consider the variation $\delta e$ :
\begin{eqnarray}
[ -\frac{(d-1)!}{4 \, d \, a_1} \, Tr(G \, h^{-1}) - 2 \, (d-2)! \, a_3 \, Tr( \mathcal{F} \, F) ] \, \delta^t_p+\nonumber \\
-2 \, [-\frac{(d-1)!}{4 \, d \, a_1} \, (G \, h^{-1})_p \, ^t -2 \, (d-2)! \,
 a_3 \, ( \mathcal{F}^{ti} \, F_{ip})]+\nonumber \\
+a_2 \, d! \, [\mbox{det} \, ( h^{-1} + \mu \mathcal{F})] ^{\alpha }\delta^t_p \, = \, 0
\label{eque e}
\end{eqnarray}
the solution is :
\begin{equation}
[-\frac{(d-1)!}{4 \, d \, a_1} \, (G \, h^{-1})_p \, ^t -2 \, (d-2)! \,
 a_3 \, ( \mathcal{F}^{ti} \, F_{ip})] \, = \, -\frac{a_2 \, d!}{d-2} \,
[\mbox{det} \, ( h^{-1} + \mu \mathcal{F})] ^{\alpha } \delta^t_p
\label{sol e}
\end{equation}
Using (\ref{def a1,a2,a3}) and (\ref{valori a2,a3}) eq. (\ref{sol e}) in matrix form  becomes :
\begin{equation}
h^{-1} \, \widehat{G} \, - \, \mathcal{F}F \, = \, \frac{1}{\alpha (d-2)} [\mbox{det} \, ( h^{-1} + \mu \mathcal{F})] ^{\alpha } 1 \! \!  1
\label{sol e matr}
\end{equation}
Now using the result:
\begin{equation}
[\mbox{det} \, (a \, \widehat{G}_{ij} + b \, F_{ij})] \, = \, [\mbox{det} \, (a \, \widehat{G}_{\mu \nu} + b \, F_{\mu \nu})] \, (\mbox{det} e)^{-2}
\label{deter}
\end{equation}
and implementing eq.(\ref{sol e matr}) for $\delta e$, we see that (\ref{eqBI 1}) becomes:
\begin{eqnarray}
(\mbox{det} \, e ) \, d^d\xi \, \frac{(d-1)!}{\alpha (d-2)} \, [\mbox{det} \, ( h^{-1} + \mu \mathcal{F})] ^{\alpha} \, = \nonumber \\
= \, (\mbox{det} \, e ) \, d^d\xi \, \frac{(d-1)!}{\alpha (d-2)}  \, [\mbox{det} \, (a \, \widehat{G}_{ij} + b \, F_{ij}) ] ^{\beta} \, = \nonumber \\
= \, (\mbox{det} \, e )^{1-2\beta}  \, d^d\xi \, \frac{ (d-1)!}{\alpha (d-2)} \,  [\mbox{det} \, (a \, \widehat{G}_{\mu \nu} + b \, F_{\mu \nu}) ] ^{\beta}
\label{finalBI}
\end{eqnarray}
Now we take $\beta \, = \, 1/2$ and so $\alpha \, = \, 1/(d-2)$. The action becomes :
\begin{equation}
S_{BI} \, =  \, (d-1)! \, \int_{M_4} \, d^d\xi \, [\mbox{det} \, (\widehat{G}_{\mu \nu} - \frac{1}{\mu} \, F_{\mu \nu}) ] ^{1/2}
\label{SBI}
\end{equation}
For $d \, = \, 4$, which is the interesting case of the $D3$--brane we obtain:
\begin{equation}
a_1 \, = \, - \, \frac{1}{8}  \hspace{1cm} a_2 \, = \, - \, \frac{1}{4} \hspace{1cm} a_3 \, = \,
\frac{3}{4}\quad ; \quad
\alpha \, \, = \, \beta \, = \, \frac{1}{2}
\label{cost d=4}
\end{equation}
\par
In this way we have shown how the kinetic part of a $Dp$--brane
action, namely the Born-Infeld type of Lagrangian can be written in
first order formalism. The new formalism can be applied to all cases
except $d=2$  where the formulae become singular. This is
just welcome since for $d=2$ we have ordinary strings for which the
Polyakov formalism is sufficient and  no world--volume cosmological
term is necessary. For $d=3$, we are instead in the case of the M2
brane or of its descendant, the D2 brane, for which no Born Infeld
action is necessary either.
\subsection{Explicit solution of the equations for the auxiliary fields for $\mathcal{F}$ and $h^{-1}$ }
In the transition to second order formalism and in the discussion of
$\kappa$-supersymmetry through the use of $1.5$ order formalism we
need the explicit solution of the first order equations and the
expression of the auxiliary fields $\mathcal{F}$, $h^{-1}$ in terms
of the physical degrees of freedom. This is what we can do most
conveniently by fixing the gauge related to the local symmetry
(\ref{transfa}) and (\ref{Picaptrans}). Our gauge choice  is provided
by setting:
\begin{equation}
  \widehat{G} = \eta
\label{gauchoice}
\end{equation}
which is identical with the yield (\ref{flattopullo}) of the $\delta
e^i$ variation in the old first order formalism. This gauge can
certainly be reached by using the degrees of freedom of
$\mathrm{GL(d,\mathbb{R})}/\mathrm{SO(1,d-1)}$. Taking
(\ref{gauchoice}) into account let us rewrite our constraint
equations into matrix form.
Eq.(\ref{sol e matr}) for the $\delta e $ variation is:
\begin{equation}
h^{-1}\widehat{G} - \mathcal{F}F \, = \, [\mbox{det} \, ( h^{-1} + \mu \mathcal{F})] ^{\alpha } 1 \! \!  1
\label{matrix e}
\end{equation}
and the other equation that we must solve is (\ref{h+fstort}) :
\begin{equation}
( h^{-1} + \mu \mathcal{F}) \, \left( \widehat{G} - \frac{1}{\mu} \, F\right) \, =  \, \left[ \mbox{det} \left(\widehat{G}  - \frac{1}{\mu} \, F \right) \right] ^{1/2}  1 \! \! 1
\label{const 2}
\end{equation}
Using our previous result for $[\mbox{det} \, ( h^{-1} + \mu \mathcal{F})] ^{\alpha
}$ we conclude that we have the following linear system of matrix
equations:
\begin{equation}
 \left\{ \begin{array}{cc}
    ( h^{-1} + \mu \mathcal{F}) \, \left( \widehat{G} - \frac{1}{\mu} \, F\right) \, =  \, \left[ \mbox{det}
  \left( \widehat{G} - \frac{1}{\mu} \, F \right) \right] ^{1/2}  1 \! \!  1  \\
    h^{-1}\widehat{G} - \mathcal{F}F \, = \, \left[ \mbox{det} \left( \widehat{G} -\frac{1}{\mu} \, F \right) \right] ^{1/2} 1 \! \!  1 \
 \end{array}\right.
\label{sistem}
\end{equation}
the solution in the gauge (\ref{gauchoice}) is:
\begin{equation}
 \left\{ \begin{array}{ccc}
 \widehat{G} \, = \, \eta \\
 \mathcal{F} \, = \frac{1}{\mu^{2}} h^{-1} F \, \eta \\
 h \eta \, = \, ( 1 \! \!  1 - \frac{1}{\mu^{2}} \, F \, \eta \, F \, \eta)
 \left[ \mbox{det} \left( \eta - \frac{1}{\mu} \, F \right) \right] ^{-1/2}
 \end{array}\right.
\label{sistem3}
\end{equation}
Since the $\eta$ metric just raises and lowers the indices  we can just ignore it and write, in more compact
form:
\begin{equation}
h  \, = \, ( \eta  - \frac{1}{\mu^{2}} \, F^{2} )
\left[ \mbox{det} \left( \eta - \frac{1}{\mu} \, F \right) \right] ^{-1/2}
\label{h no eta}
\end{equation}
\section{The $D3$--brane example and $\kappa$-supersymmetry}
\label{thed3examp}
In this section we focus on the case $d=4$ and we apply our new first
order formalism to the description of the $\kappa$-supersymmetric
action of a $D3$--brane. As claimed in the introduction,
$\kappa$--supersymmetry just follows, via a suitable projection, from
the bulk supersymmetries as derived from supergravity, the type II B
theory, in this case. The latter has a duality symmetry with respect
to an $\mathrm{SL(2,\mathbb{R})}$ group of transformations that acts
non linearly on the two scalars of massless spectrum, the dilaton $\phi$ and
the Ramond scalar $C_0$. Indeed these two parametrize the coset
manifold $\mathrm{SL(2,\mathbb{R})}/\mathrm{O(2)}$ and actually
correspond to its solvable parametrization (see eq.(\ref{Lfatcoset}) of the
appendix).
Hence the $D3$--brane action we want to write, not only should be cast into
first order formalism, but should also display manifest covariance
with respect to $\mathrm{SL(2,\mathbb{R})}$. This covariance relies
on introducing a two component charge vector $q_\alpha$ that transforms in the
fundamental representation of $\mathrm{SU(1,1)}$
and expresses the charges carried by the $D3$ brane with respect to
the $2$--forms $A^\alpha_{[2]}$ of bulk supergravity (both the Neveu Schwarz $B_{[2]}$ and
Ramond--Ramond $C_{[2]}$). According to the geometrical formulation of type IIB
supergravity which is summarized in the appendix we set:
\begin{equation}
\begin{array}{rclcrcl}
A^{\Lambda} & = & (B_{[2]} \, , \, C_{[2]}) &;&
A^{\alpha} & =& \mathcal{C}^{\alpha} \, \, _{\Lambda} \, A^{\Lambda}
\\
A^{\alpha  = 1} & = & \frac{1}{\sqrt{2}} \, (B_{[2]} \, - \, i \, C_{[2]})
&;&
A^{\alpha  = 2} & = & \frac{1}{\sqrt{2}} \, (B_{[2]} \, + \, i \,
C_{[2]})\
\end{array}
\label{def A}
\end{equation}
and by definition we call $\epsilon_{\alpha\beta} \,q^{\beta}$ the orthogonal complement of $q_\alpha$:
\begin{equation}
  q_\alpha \, q^\alpha =1 \quad ; \quad q_\alpha \, q_\beta \,
  \epsilon^{\alpha\beta}=0
\label{qvecti}
\end{equation}
In terms of these objects we write down the complete action of the $D3$--brane as follows:
\begin{eqnarray}
\mathcal{L} & = & \Pi^{\underline{a}} _i \, V^{\underline{b}} \, \eta_{\underline{ab}}
 \, \eta^{i\ell_1} \, \wedge e^{\ell_2} \, \wedge \, \dots \wedge e^{\ell_4} \, \epsilon_{\ell_1 \dots \ell_4}
  + a_1 \, \Pi^{\underline{a}} _i \, \Pi^{\underline{b}}
_j \, \eta_{\underline{ab}} \, h^{ij} \, e^{\ell_1} \, \wedge \, \dots
\wedge e^{\ell_4} \, \epsilon_{\ell_1 \dots \ell_4} \nonumber\\
&&+ a_2 \, \left [ \mbox{det}\,\left(  h^{-1} + \mu \mathcal{F}\right) \right] ^\alpha \,
  e^{\ell_1} \, \wedge \, \dots
\wedge e^{\ell_4} \, \epsilon_{\ell_1 \dots \ell_4}\nonumber\\
&& + a_3 \mathcal{F}^{ij} \, F^{[2]} \, \wedge \, e^{\ell_3} \, \wedge \,e^{\ell_4} \,
\epsilon_{ij\ell_3 \ell_4} \nonumber \\
&&+\nu \, F \, \wedge \, F -i \, a_5 q^{\alpha} \, \epsilon_{\alpha \beta} \, A^{\beta} \,
\wedge \, F + a_6 \, C_{[4]}
\label{us eq}
\end{eqnarray}
where $C_{[4]}$ is the $4$--form potential, the coefficients
\begin{equation}
\alpha\, = \, \frac{1}{2} \hspace{1cm} a_1 \, = \, - \, \frac{1}{8}  \hspace{1cm} a_2 \, = \, - \,
\frac{1}{4} \hspace{1cm}
  a_3\, = \, \frac{3}{4}
\label{repetita}
\end{equation}
have already been determined, while $a_5,a_6,\nu$ are new
coefficients to be fixed by $\kappa$--supersymmetry. The first two
are numerical, while $\nu$ will also depend on the bulk scalars.
In the action (\ref{us eq})
\begin{equation}
  F^{[2]} \equiv dA^{[1]} + q_{\alpha} A^{\alpha}
\label{nFdefidA}
\end{equation}
is the field strength of the world--volume gauge field and depends on
the charge vector $q^{\alpha}$. The physical interpretation of
$F^{[2]}$ is as follows. By definition a $Dp$--brane is a locus in
space--time where open strings can end or, in the dual picture,
boundaries for closed string world--volumes can be located. The type IIB theory contains
two kind of strings, the fundamental strings and the $D$-strings
which are rotated one into the other by the $\mathrm{SL(2,\mathbb{Z})}\subset
\mathrm{SL(2,\mathbb{R})} $ group. Correspondingly a $D3$ brane can be
a boundary either for fundamental or for $D$--strings or for a
mixture of the two. The charge vector $q^\alpha$ just expresses this
fact and characterizes the $D3$--brane as a boundary for
strings of $q$--type. Furthermore the definition (\ref{nFdefidA}) of $F^{[2]}$
encodes the following idea: the world--volume gauge $1$-form $A^{[1]}$
is just the parameter of a gauge transformation for the $2$--form $q_{\alpha}
A^{\alpha}$, which in a space--time with boundaries can be reabsorbed
everywhere except on the boundary itself. Note that if we
take $q_{\alpha}  = \frac{1}{\sqrt{2}}(1 \, , \, 1)$ we obtain :
\begin{equation}
q_{\alpha} A^{\alpha} \, = \, B_{[2]} \quad ; \quad
 -i \,  q^{\alpha} \, \epsilon_{\alpha \beta} \, A^{\beta} \, = \, C_{[2]}
\label{no q}
\end{equation}
\subsection{$\kappa$--supersymmetry}
Next we want to prove that with an appropriate choice of $\nu, a_5$
and $a_6$ the action (\ref{us eq}) is invariant against bulk
supersymmetries characterized by a projected spinor parameter. For
simplicity we do this in the case of the choice $q_{\alpha}  = \frac{1}{\sqrt{2}}(1 \, , \,
1)$. For other choices of the charge type the modifications needed in
the prove will be obvious from its details.
\par
To accomplish our goal we begin by writing the supersymmetry transformations of the
bulk differential forms $V^{\underline{a}}$, $B_{[2]}$, $C_{[2]}$ and
$C_{[4]}$ which appear in the action. From the rheonomic
parametrizations (\ref{tors1},\ref{lore1},\ref{2form1},\ref{4form1}) we immediately
obtain:
\begin{eqnarray}
&&\delta V^{\underline{a}}\, = \, \mbox{i}\, \ft 1 2 \, \left( \overline{\epsilon} \,
\Gamma^{\underline{a}} \, \psi + \overline{\epsilon}^{*} \,
\Gamma^{\underline{a}} \, \psi^{*}\right)   \nonumber\\
&&\delta B_{[2]} \, = \, - \, 2 \, i \, [(\Lambda^{1}_{+} +\Lambda^{2}_{+}) \,
{\bar\epsilon} \, \Gamma_{\underline{a}} \, \psi^{*} \,
V^{\underline{a}}+(\Lambda^{1}_{-} +\Lambda^{2}_{-}) \, {\bar\epsilon}^{*} \,
\Gamma_{\underline{a}} \, \psi \, V^{\underline{a}}] \nonumber\\
&&\delta C_{[2]} \, = \, 2 \, [(\Lambda^{1}_{+} -\Lambda^{2}_{+}) \, {\bar\epsilon} \,
\Gamma_{\underline{a}} \, \psi^{*} \,
V^{\underline{a}}+(\Lambda^{1}_{-} -\Lambda^{2}_{-}) \, {\bar\epsilon}^{*} \,
\Gamma_{\underline{a}} \, \psi \, V^{\underline{a}}] \nonumber\\
&&\delta C_{[4]} \, = \underbrace{ -\frac{1}{6} ( {\bar\epsilon} \, \Gamma_{\underline{abc}}
\, \psi \, - \, {\bar\epsilon}^{*} \, \Gamma_{\underline{abc}} \, \psi^{*})
\, V^{\underline{abc}} }_{\delta C_{[4]}^{\prime}} + \frac{1}{8} \,  [ B_{[2]} \,
\delta C_{[2]} \, - \, C_{[2]} \, \delta B_{[2]} ]
\label{susy tran}
\end{eqnarray}
Note that in writing the above transformations we have neglected all
terms involving the dilatino field. This is appropriate since the
background value of all fermion fields is zero. The gravitino
$1$--form $\psi$ is instead what we need to keep track of. Proving
$\kappa$--supersymmetry is identical with showing that all $\psi$
terms cancel against each other in the variation of the action.
Relying on (\ref{susy tran}) the variation of the W.Z.T term is as follows:
\begin{eqnarray}
&&\delta ( \nu \, F \wedge \, F +a_5 \, C_{[2]} \wedge F + C_{[4]}) \, = \nonumber\\
&&= 2 \, \nu \, B \, \delta B + a_5  \, B \, \delta C_{[2]} + \frac{1}{8} \, a_6 \, B \, \delta C + a_5 \, C \,
\delta B -\frac{1}{8} \, a_6 \, C \, \delta B +
 a_6 \, \delta C_{[4]}^{\prime}
\label{var wzt}
\end{eqnarray}
if we set $ a_6 \, = \, 8 \, a_5$ the variation (\ref{var wzt}) simplifies to:
\begin{equation}
\delta(W.Z.T) \, = \, 2 \, B \, (\nu \, \delta B + a_5 \, \delta C) + 8 \, a_5 \, \delta C_{[4]}^{\prime}
\label{var 2  wzt}
\end{equation}
and with such a choice the complete variation of the Lagrangian under a supersymmetry  transformation of arbitrary
parameter is:
\begin{eqnarray}
\delta \mathcal{L} \, & = & \, \delta \mathcal{L}_{\psi} +  \delta \mathcal{L}_{\psi^{*}} \nonumber\\
\delta \mathcal{L}_{\psi} \, & = & \, [- \, 3! \, i \, \Pi^{\underline{a},p} \, ( {\bar\epsilon} \, \Gamma^{\underline{b}} \, \psi ) \eta_{\underline{ab}} \, + \,
(\mu_{1} \, \mathcal{F}^{ip} \, + \, \mu_{2} \, {\tilde{F}}^{ip}) \, V^{\underline{a}}_{i} \, ( {\bar\epsilon}^{*} \, \Gamma_{\underline{a}} \, \psi ) \, + \, \nonumber\\
&&-\frac{4}{3} \, a_5 \, ( {\bar\epsilon} \, \Gamma_{\underline{abc}} \, \psi ) \,  V^{\underline{a}}_{i} \, V^{\underline{b}}_{j} \, V^{\underline{c}}_{k} \,
\epsilon^{ijkp}] \, \Omega^{[3]}_{p} \nonumber\\
\delta \mathcal{L}_{\psi_{*}} \, & = & \, [ - \, 3! \, i \, \Pi^{\underline{a},p} \, ( {\bar\epsilon}^{*} \, \Gamma^{\underline{b}} \, \psi^{*} ) \eta_{\underline{ab}} \, + \,
(\mu_{3} \, \mathcal{F}^{ip} \, + \, \mu_{4} \, {\tilde{F}}^{ip}) \, V^{\underline{a}}_{i} \, ( {\bar\epsilon} \, \Gamma_{\underline{a}} \, \psi^{*} ) \, + \, \nonumber\\
&&+\frac{4}{3} \, a_5 \, ( {\bar\epsilon}^{*} \, \Gamma_{\underline{abc}} \, \psi^{*} ) \,  V^{\underline{a}}_{i} \, V^{\underline{b}}_{j} \, V^{\underline{c}}_{k} \, \epsilon^{ijkp}] \, \Omega^{[3]}_{p}
\label{delta L}
\end{eqnarray}
where :
\begin{equation}
\begin{array}{rclcrcl}
\mu_{1} \, & = & \, - \, 8 \, i \, a_3 \, (\Lambda^{1}_{-} \, + \, \Lambda^{2}_{-})
&;&
\mu_{2} \, & = & \, 8 \, [- i \, \nu \,  (\Lambda^{1}_{-} \, + \, \Lambda^{2}_{-}) \, + \, a_5  \,
 (\Lambda^{1}_{-} \, - \, \Lambda^{2}_{-})] \\
\mu_{3} \, & = & \, - \, 8 \, i \, a_3 \, (\Lambda^{1}_{+} \, + \, \Lambda^{2}_{+})
&;&
\mu_{4} \, & = & \, 8 \, [- i \, \nu \,  (\Lambda^{1}_{+} \, + \, \Lambda^{2}_{+}) \, + \, a_5
\, (\Lambda^{1}_{+} \, - \, \Lambda^{2}_{+})] \
\end{array}
\label{def mu}
\end{equation}
Recalling eq.s (\ref{Lambigcos}) and (\ref{Lfatcoset}) of the appendix
the above eq.s (\ref{def mu}) become:
\begin{equation}
\begin{array}{rclcrcl}
\mu_{1} \, & = & \, -6 \, i \, e^{\phi/2} &;&
\mu_{2} \, & = & \, 8 \, a_5  e^{-\frac{\phi}{2}} \\
\mu_{3} \, & = & \, -6 \, i \,  \, e^{\phi/2} &;&
\mu_{4} \, & = & \, - \, 8 \, a_5  e^{-\frac{\phi}{2}}\
\end{array}
\label{def mu cay}
\end{equation}
where we have chosen:
\begin{eqnarray}
\nu & = & - \, a_5 \, C_{0} \,= \, a_5 \, \mbox{Re} \, \mathcal{N}
\label{val nu}
\end{eqnarray}
In the above equation we have introduced the complex kinetic matrix which would appear in a
gauge theory with scalars sitting in $\mathrm{SU(1,1)/U(1)}$ and
determined by the classical Gaillard--Zumino general formula \footnote{For a general discussion of the
Gaillard-Zumino formula see for instance \cite{parilez} } applied
to the specific coset:
\begin{equation}
  \mathcal{N} = \mbox{i}\,\frac{\Lambda^1_- - \Lambda^2_-}{\Lambda^1_- +
  \Lambda^2_-} \quad \Rightarrow \quad \cases{ \mbox{Re} \,
  \mathcal{N} = - C_0 \cr
  \mbox{Im} \, \mathcal{N} = e^{-\phi}\cr}
\label{scriptaNmata}
\end{equation}
It is convenient to rewrite the full variation (\ref{delta L}) of the Lagrangian in matrix form in the
$2$--dimensional space spanned by the fermion parameters $\left( \epsilon \, , \, \epsilon ^{*}\right) $ :
\begin{equation}
\delta \mathcal{L} \, = \, \delta \mathcal{L}_{\psi} +  \delta \mathcal{L}_{\psi^{*}} \,
= \, ( {\bar\epsilon} \, , \,  {\bar\epsilon}^{*})  \,
A  \, \,  \left( \begin{array}{c}
     \psi \\
    \psi^{*} \
 \end{array}\right)
\label{matr form}
\end{equation}
\begin{equation}
A_{k} \, = \, \left( \begin{array}{ccc}
     - 6 \, i \, \gamma_{k} -\frac{4}{3} \, a_5 \, \gamma_{ijl} \, \epsilon^{ijlm} \, h_{mk}  &,&
       (\mu_{3} \, \mathcal{F}^{lm} \, + \, \mu_{4} \, {\tilde{F}}^{lm}) \, h_{mk} \, \gamma_{l}\\
      (\mu_{1} \, \mathcal{F}^{lm} \, + \, \mu_{2} \, {\tilde{F}}^{lm}) \, h_{mk} \, \gamma_{l} & , &
       - 6 \, i \, \gamma_{k} +\frac{4}{3} \, a_5 \, \gamma_{ijl} \, \epsilon^{ijlm} \, h_{mk}\
 \end{array}\right)
\label{matrix A}
\end{equation}
where $A \, = \, A_{k} \, \Omega_{[3]}^{k} $, and  $\Omega_{[3]}^{k}\equiv \eta^{k\ell} \,
\epsilon_{\ell ijk} \, e^i \wedge e^j \wedge e^k.$ denotes the quadruplet of three--volume forms\\
The matrix $A_{k}$ is a tensor product of a matrices in spinor space and $2\times 2$ matrices
in the space spanned by $\left( \epsilon \, , \, \epsilon ^{*}\right)
$. It is convenient to spell out this tensor product structure which
is achieved by the following rewriting:
\begin{equation}
A_{k} = f_{1} \, \gamma_{k} \otimes 1 \! \! 1 + f_{2} \, \tilde{\gamma} \,
 ^{m} h_{mk} \otimes \sigma_{3} + f_{3} \, \Pi^{m}_{1} h_{mk} \otimes \sigma_{1} + f_{4} \, \Pi^{m}_{2} h_{mk}
 \otimes \sigma_{2}
\label{concis Ak}
\end{equation}
where :
\begin{equation}
\begin{array}{rclcrclcrclcrcl}
f_{1} & = & -6 i &;& f_{3}& =& -6 i   &;&
f_{2} & = & - \frac{4}{3} a_5  &;& f_{4} &=& -8i a_{5}
\end{array}
\label{gli f}
\end{equation}
and:
\begin{equation}
\begin{array}{rclcrclcrcl}
\tilde{\gamma} \, ^{m} & \equiv & \gamma_{ijl} \, \epsilon^{ijlm}
&;&
\Pi^{m}_{1} & \equiv & e^{\phi /2} \mathcal{F}^{lm} \gamma_{l} &;&
\Pi^{m}_{2} & \equiv & e^{ - \phi /2} \tilde{F}^{lm} \gamma_{l}
\end{array}
\label{pi 12}
\end{equation}
now using (\ref{sistem3}), (\ref{h no eta})  we set
\begin{eqnarray}
 \frac{1}{\mu} & = & e^{-\phi/2} = \sqrt{\mbox{Im} \,
 \mathcal{N}}\nonumber\\
 \hat{F}  & \equiv & \sqrt{\mbox{Im} \,
 \mathcal{N}} F \nonumber \\
\label{muchoi}
\end{eqnarray}
and we obtain:
\begin{eqnarray}
&&\Pi^{m}_{1} h_{mk}  =  e^{\phi /2} \mathcal{F}^{lm} \gamma_{l} h_{mk} =
e^{\phi /2} e^{ - \phi}(Fh^{-1})^{lm} h_{mk} \gamma_{l} \equiv \hat{F}_{lk} \gamma^{l} \equiv \Pi_{k} \nonumber \\
&&\Pi^{m}_{2}  \equiv  e^{ - \phi /2} \tilde{F}^{lm} \gamma_{l} \equiv
 \tilde{\hat{F}} \, ^{lm} \gamma_{l} \equiv \tilde{\Pi} \, ^{m}
\label{pi pi tilda}
\end{eqnarray}
This observation further simplifies the expression of $A_{k}$ which can be rewritten as:
\begin{equation}
A_{k} = f_{1} \, \gamma_{k} \otimes 1 \! \! 1 + f_{2} \, \tilde{\gamma} \, ^{m} h_{mk} \otimes \sigma_{3}
+ f_{3} \, \Pi^{k} \otimes \sigma_{1} + f_{4} \, \tilde{\Pi} \, ^{m} h_{mk} \otimes \sigma_{2}
\label{concis2 Ak}
\end{equation}
The proof of $\kappa$--supersymmetry can now be reduced to the
following simple computation. Assume we have a matrix operator $\Gamma$
with the following properties:
\begin{eqnarray}
  \mbox{[a]}&\,& \Gamma^{2} = 1 \! \!  1 \nonumber\\
  \mbox{[b]}&\,& \Gamma \, A_{k} \, = \,  A_{k}
  \label{properte}
\end{eqnarray}
It follows that
\begin{equation}
P \, = \, \frac{1}{2} ( 1 \! \!  1 \, - \, \Gamma)
\label{proyector}
\end{equation}
is a projector since $P^2 =1 \! \!  1$ and that
\begin{equation}
P A_{k} = \frac{1}{N}(1 \! \! 1 - \Gamma) \, A_{k} = 0
\label{k supersi pro}
\end{equation}
Therefore if we use supersymmetry parameters $\left( \overline{\kappa} , \overline{\kappa}^*\right)  =
\left (\overline{\epsilon} ,
\overline{\epsilon} ^*\right) \, P$ projected with this $P$, then the
action is invariant and this is just the proof of
$\kappa$--supersymmetry.
\par
The appropriate $ \Gamma $ is the following \cite{project}
\footnote{In the paper quoted above the $\kappa$--supersymmetry projector presented here was originally introduced
within a 2nd order formulation of the theory. It is particularly significant and rewarding that the same
projector is valid also in first order formulation. As shown in the appendix the mechanism by means of which it works
are very subtle and take advantage of the explicit solutions for the auxiliary fields in terms of the physical ones.
In this way one finds an overall non trivial check of all the algebraic machinery of our new first order formalism.}:
\begin{equation}
\Gamma \, = \, \frac{1}{N} [(\omega_{[4]} \, + \, \omega_{[0]}) \, \otimes \,
\sigma_{3} \, + \, \omega_{[2]} \, \otimes \, \sigma_{2}]
\label{def gamma}
\end{equation}
where:
\begin{eqnarray}
\omega_{[4]} \, & = & \, \alpha_{4} \, \epsilon^{ijkl} \, \gamma_{ijkl} \nonumber \\
\omega_{[0]} \, & = & \, \alpha_{0} \,  \epsilon^{ijkl} \, \hat{F}_{ij} \, \hat{F}_{kl} \nonumber \\
\omega_{[2]} \, & = & \, \alpha_{2} \,\epsilon^{ijkl} \, \hat{F}_{ij} \,
\gamma_{kl}\nonumber\\
N \,& =& \, \left[ \mbox{det} \left( \eta \, \pm \, \hat{F} \right) \right] ^{1/2}
\label{def omega}
\end{eqnarray}
and the coefficients are fixed to:
\begin{equation}
\begin{array}{rclcrclcrcl}
\alpha_4 & = & \frac{1}{24} &;&
\alpha_0 & = & \frac{1}{8} &;&
\alpha_2 & = & \frac{i}{4}\
\end{array}
\label{def alfa}
\end{equation}
This choice suffices to guarantee property $\mbox{[a]}$ in the above
list. Property $[b]$ is also verified if one chooses:
\begin{equation}
  a_{5}= \frac{3}{4}\mbox{i}
\label{a5value}
\end{equation}
The proof of the two properties is given in the appendix
\ref{projectap}. Essential ingredients in that proof are the
following identities holding true for any antisymmetric tensor
$\widehat{F}$:
\begin{equation}
\mbox{det} \left( \eta \, \pm \, \hat{F} \right) = - \,1 \, + \, \frac{1}{2} Tr(\hat{F}^{2}) \,
+ \, \left( \frac{1}{8} \epsilon^{ijkl} \, \hat{F}_{ij} \, \hat{F}_{kl} \right)^2
\label{determinante eta}
\end{equation}
and
\begin{eqnarray}
\hat{F} \, \tilde{\hat{F}} & = & - \,\frac{1}{8}\left( F_{ij} \, F_{kl} \, \epsilon^{ijkl} \right)
\, 1 \! \! 1 \, = \, - \omega_{[0]} \, 1 \! \! 1 \nonumber \\
\hat{F}^{2} + \tilde{\hat{F}}^{2} & = &\frac{1}{2} \, Tr(F^{2}) \,  1 \! \! 1
\label{prop 2 3}
\end{eqnarray}
\section{Outlook and conclusions}
\label{outlook}
In this paper we have introduced a new first order formalism for
$p$--brane world volume actions that allows to reproduce the
Born--Infeld second order action via the elimination of a set composed by three auxiliary fields:
\begin{itemize}
  \item $\Pi^{\underline{a}}_i$
  \item $h^{ij}$ (symmetric)
  \item $\mathcal{F}^{ij}$ (antisymmetric)
\end{itemize}
Distinctive properties of our new formulation are:
\begin{enumerate}
  \item All fermion fields are implicitly hidden inside
  the definition of the $p$--form potentials of supergravity
  \item $\kappa$--supersymmetry is easily proven from supergravity
  rheonomic parametrization
  \item The action is manifestly covariant with respect to the duality group
  $\mathrm{SL(2,\mathbb{R})}$ of type IIB supergravity.
  \item The action functional can be computed on any background which
  is an exact solution of the supergravity bulk equations.
\end{enumerate}
Of specific interest in applications are precisely the last two
properties. Putting together our result we can summarize the $D3$ brane action we have found as
follows:
\begin{eqnarray}
\mathcal{L} & = & \Pi^{\underline{a}} _i \, V^{\underline{b}} \, \eta_{\underline{ab}}
 \, \eta^{i\ell_1} \, \wedge e^{\ell_2} \, \wedge \, \dots \wedge e^{\ell_4} \, \epsilon_{\ell_1 \dots \ell_4}
  - \ft 1 8 \, \Pi^{\underline{a}} _i \, \Pi^{\underline{b}}
_j \, \eta_{\underline{ab}} \, h^{ij} \, e^{\ell_1} \, \wedge \, \dots
\wedge e^{\ell_4} \, \epsilon_{\ell_1 \dots \ell_4} \nonumber\\
&&- \ft 1 4 \, \left [ \mbox{det}\,\left(  h^{-1} + \sqrt{\mbox{Im}\mathcal{N}} \mathcal{F}\right) \right] ^{1/2} \,
  e^{\ell_1} \, \wedge \, \dots
\wedge e^{\ell_4} \, \epsilon_{\ell_1 \dots \ell_4}\nonumber\\
&& + \ft 3 4 \, \mathcal{F}^{ij} \, F \, \wedge \, e^{\ell_3} \, \wedge \,e^{\ell_4} \,
\epsilon_{ij\ell_3 \ell_4} \nonumber \\
&& + \ft 3 4 i \, \mbox{Re}\,\mathcal{N}\, \mbox{}\, F \, \wedge \, F + \, \ft 3 4 \, q^{\alpha} \,
\epsilon_{\alpha \beta} \, A^{\beta} \,
\wedge \, F + 6 i \, C_{[4]}\nonumber\\
 F & \equiv & dA^{[1]} + q_{\alpha} A^{\alpha}\nonumber\\
\mathcal{N} &=& \mbox{i}\,\frac{\Lambda^1_- - \Lambda^2_-}{\Lambda^1_- +
  \Lambda^2_-}
\label{D3branaczia}
\end{eqnarray}
Evaluating for instance the above action  on the background provided
by the bulk solution found in \cite{noialtrilast} which describes a $D3$--brane with an
$\mathbb{R}^2 \times ALE$ transverse manifold we can finally write the appropriate
source term of that exact solution
which was so far missing. Alternatively by expanding (\ref{D3branaczia}) for small fluctuations around
the same background we can use it as a token to explore
the gauge/gravity correspondence.
\par
In relation with the solution \cite{noialtrilast} which is one of the
main motivations for undertaking the present investigation we note that
the $D3$--brane action (\ref{D3branaczia}) is not sufficient to work
out all the sources for such a solution. Indeed (\ref{D3branaczia})
gives account of the $C_{[4]}$ charge but not of the $B_{[2]}$ or
$C_{[2]}$ charges. The latter are however essential in the solution
\cite{noialtrilast} since there we also have
nontrivial $2$--forms. It follows that to complete our task we
need to adjoin to (\ref{D3branaczia}) also a source action for the
$2$--forms which cannot be anything else but a $5$--brane, since
the $2$--forms are magnetically coupled. A distinctive feature of
the needed $5$--brane is that it should not couple to the dilaton
field since the latter is constant in the solution \cite{noialtrilast}. For a
Neveu-Schwarz $5$--brane or for a $D5$--brane this is impossible yet
it can be possible for a mixture of the two. This singles out an
obvious research direction which we are presently pursuing. That is
applying our new first order formalism to the case of a $q$--type $5$--brane.
\par
Several other applications of our formalism are possible since it
provides a general way to deal with Born--Infeld,
$\kappa$--supersymmetric actions. In particular it can give new
insight on the open problem of writing supersymmetric, non abelian,
Born Infeld actions.
\medskip
\section*{Acknowledgements}
We are grateful to Marco Bill\'o, Igor Pesando, Carl Hermann, Mario
Trigiante and Matteo Bertolini for many important and clarifying discussions at thebeginning of this work.
\newpage
\appendix
\section{Summary of type IIB supergravity in a geometrical set up}
\label{type2bsum}
The formulation of type IIB supergravity as it appears in string
theory textbooks \cite{greenschwarz,polchinski} is tailored for the
comparison with superstring amplitudes and is quite appropriate to this goal. Yet, from the
viewpoint of the general geometrical set up of supergravity theories
this formulation is somewhat unwieldy. Specifically it  neither makes the $\mathrm{SU(1,1)/U(1)}$
coset structure of the theory
manifest, nor it  relates the supersymmetry transformation rules  to the
underlying algebraic structure which,  as in all other instances of
supergravities, is a simple and well defined {\sl Free Differential
algebra}.
\par
The Free Differential Algebra of type IIB supergravity was singled out
many years ago by Castellani in \cite{castella2b} and the geometric
manifestly $\mathrm{SU(1,1)}$ covariant formulation of the theory was
constructed by Castellani and Pesando in \cite{igorleo}. In this
appendix we summarize their formulae giving also their
transcription from a complex $\mathrm{SU(1,1)}$ basis to a real
$\mathrm{SL(2,\mathbb{R})}$ basis. Furthermore we provide the
translation vocabulary between these intrinsic notations and those
of Polchinski's textbook \cite{polchinski} frequently used in current superstring
literature.
\subsection{The $\mathrm{SU(1,1)/U(1)}\sim\mathrm{ SL(2,\mathbb{R})/O(2)}$ coset}
\label{sl2rcoset}
The basic ingredient in all supergravity constructions is the
parametrization of the scalar manifold geometry  that, with few
exceptions, corresponds to a homogeneous scalar manifold \cite{parilez}. In all
these cases the essential building block appearing in the Lagrangian
and supersymmetry transformation rules is the coset representative
$\mathbb{L}(\phi_i)$ that provides a parametrization of the coset
manifold $G/H$ in terms of some chosen patch of coordinates. A very
useful choice is given by the so called solvable Lie algebra
parametrization \footnote{For a review see either \cite{mariothesis}
or \cite{parilez} and all references therein}. This is true also in
the present case where the solvable parametrization of the coset
$\mathrm{SU(1,1)/U(1)}\sim\mathrm{ SL(2,\mathbb{R})/O(2)}$ is
precisely that which allows for the  identification of the massless
superstring fields inside the covariant formulation of supergravity.
\par
Our notations are as follows.
\par
\leftline{\underline{\sl $\mathrm{SL(2,\mathbb{R})}$ Lie algebra}}
\begin{equation}
  \left[ L_0 \, , \, L_\pm \right] = \pm \, L_\pm \quad ; \quad \left[ L_+ \, , \, L_- \right] = 2 \,
  L_0
\label{sl2alg}
\end{equation}
with explicit $2$--dimensional representation:
\begin{equation}
  L_0= \ft 1 2 \, \left( \begin{array}{cc}
    1 & 0 \\
    0 & -1 \
  \end{array}\right) \quad ; \quad L_+= \left( \begin{array}{cc}
    0 & 1 \\
    0 & 0 \
  \end{array}\right) \quad ; \quad L_-= \left( \begin{array}{cc}
    0 & 0 \\
    1 & 0 \
  \end{array}\right) \quad ; \quad
\label{2drepsl2r}
\end{equation}
\par
\leftline{\underline{\sl Coset representative of $\mathrm{SL(2,\mathbb{R})/O(2)}$ in the solvable parametrization}}
\begin{equation}
  \mathbb{L}\left( \varphi , C_{[0]} \right) =\exp \left[  \varphi \, L_0 \right]  \, \exp \left[
  C_{[0]} e^{\varphi} \, L_-\right] \, = \, \left( \begin{array}{cc}
    \exp[\varphi/2] & 0 \\
    C_{[0]}e^{\varphi/2} & \exp[-\varphi/2] \
  \end{array}\right)
\label{Lfatcoset}
\end{equation}
where $\varphi(x)$ and $ C_{[0]}$ are respectively identified with
the dilaton and with the Ramond-Ramond 0-form of the superstring
massless spectrum.
The isomorphism of $\mathrm{SL(2,\mathbb{R})}$ with
$\mathrm{SU(1,1)}$ is realized by conjugation with the Cayley matrix:
\begin{equation}
  \mathcal{C}= \ft {1}{\sqrt{2}} \, \left( \begin{array}{cc}
    1 & - {\rm i} \\
    1 & {\rm i} \
  \end{array}\right)
\label{caylmat}
\end{equation}
Introducing the $\mathrm{SU(1,1)}$ coset representative
\begin{equation}
\mathrm{SU(1,1)}\,  \ni \,\Lambda \, = \, \mathcal{C} \, \mathbb{L}
\, \mathcal{C}^{-1}
\label{Lambigcos}
\end{equation}
from the left invariant $1$--form $\Lambda^{-1} \, d \Lambda $ we can
extract the $1$-forms corresponding to the scalar vielbein $P$ and the $\mathrm{U(1)}$ connection $Q$
\par
\leftline{\underline{\sl The $\mathrm{SU(1,1)/U(1)}$ vielbein and
connection}}
\begin{equation}
  \Lambda^{-1} \, d \Lambda \, = \, \left( \begin{array}{cc}
   - {\rm i}\, Q & P \\
    P^\star & {\rm i}\, Q \
  \end{array} \right)
\label{su11viel}
\end{equation}
Explicitly
\begin{equation}
\begin{array}{rcll}
P & = & \ft 1 2 \, \left( d\varphi - {\rm i}\, e^\varphi\, dC_{[0]} \right)  & \mbox{scalar vielbein} \\
Q & = & \ft 1 2 \,  \exp[ \varphi ] \, dC_{[0]} & \mbox{$\mathrm{U(1)}$-connection} \
\end{array}
\label{PQvalue}
\end{equation}
\subsection{The free differential algebra, the supergravity fields and the curvatures}
Following Castellani and Pesando the field content of type IIB
supergravity is organized into representations of $\mathrm{SU(1,1)}$
as displayed in table \ref{tabella}.
\begin{table}
  \centering
  \caption{\textbf{Field content of type IIB supergravity}.
 {\sl The early Greek indices $\alpha,\beta,\dots=1,2$ run in the fundamental representation of $\mathrm{SU(1,1)}$,
 while the early capital Latin indices $A,B,\dots =1,2$ run in the fundamental representation of
 $\mathrm{SL(2,\mathbb{R})}$. The $p$-gauge forms of the Ramond Ramond sector are denoted by $C_{[p]}$. }
 \label{tabella}}
$$
  \vbox{
  \offinterlineskip
  \halign
    {&\vrule# &\strut\hfil #\hfil &\vrule# &\hfil #\hfil &\vrule#
     &\hfil #\hfil &\vrule# &\hfil #\hfil  &\vrule# \cr
   \noalign{\hrule}
      height5pt
      &\omit& &\omit& &\omit& &\omit&
      \cr
      &\hbox to4 true cm{\hfill Field in $\mathrm{SU(1,1)}$ basis \hfill}&
      &\hbox to4 true cm{\hfill SU(1,1) repres. \hfill}&
      &\hbox to4 true cm{\hfill U(1) charge \hfill}&
      &\hbox to4 true cm{\hfill \vbox {\hbox{superstring}
       \hbox{zero modes}} \hfill}&
      \cr
      height5pt
      &\omit& &\omit& &\omit& &\omit&
      \cr
    \noalign{\hrule}
      height5pt
      &\omit& &\omit& &\omit& &\omit&
      \cr
      &$V^a_\mu$& &$J=0$& &0& &\mbox{graviton $h_{\mu\nu}$}& \cr
      height10pt
      &\omit& &\omit& &\omit& &\omit&
      \cr
      &$\psi_\mu$& &J=0& &${1\over2}$& &$\mbox{ gravitinos $\psi_{A\mu}$}$  & \cr
      height10pt
      &\omit& &\omit& &\omit& &\omit&
      \cr
      &$A^\alpha_{\mu\nu}$& &$J=\frac{1}{2}$& &0& & $B_{[2]} \, , \, C_{[2]}  $ &\cr
      height10pt
      &\omit& &\omit& &\omit& &\omit&
      \cr
      &$C_{\mu\nu\rho\sigma}$& &$J=0$& &0& &$C_{[4]}$& \cr
      height10pt
      &\omit& &\omit& &\omit& &\omit&
      \cr
      &$\lambda$& &$J=0$& &${3\over2}$& &$\mbox{dilatinos $\lambda_A$}$& \cr
      height10pt
      &\omit& &\omit& &\omit& &\omit&
      \cr
      &$\mathbb{L}^\alpha_{\phantom{\alpha}\beta}$& &$J=\frac{1}{2}$& &$\pm 1$& &$\varphi, C_{[0]}$& \cr
      height5pt
      &\omit& &\omit& &\omit& &\omit&
      \cr
    \noalign{\hrule} }
  }
$$
  \label{fieldcont}
\end{table}
In order to write down the free differential algebra the critical
issue is the correct identification of the fermionic terms contributing  to the curvature
of the complex $2$-form  doublet
$A^\alpha_{\mu\nu}$. These latter transform in the $2$--dimensional
representation of $\mathrm{SU(1,1)}$ and  are related by the Cayley
matrix of eq.(\ref{caylmat}) to a doublet of real $2$-forms $\mathbf{A}^\Lambda_{\mu\nu}$
that transform in the $2$--dimensional representation of
$\mathrm{SL(2,\mathbb{R})}$:
\begin{equation}
 \left(  \begin{array}{c}
   A^1_{\underline{\mu\nu}} \\
   A^2_{\underline{\mu\nu}}  \
 \end{array}\right)  = \mathcal{C} \left(  \begin{array}{c}
   \mathbf{A}^1_{\underline{\mu\nu}} \\
   \mathbf{A}^2_{\underline{\mu\nu}}  \
 \end{array}\right)
\label{transforA}
\end{equation}
We introduce a doublet of Majorana-Weyl spinor 1-forms (the
gravitinos) having the same chirality:
\begin{equation}
  \Gamma_{11} \, \psi_A = -\psi_{A} \quad ; \quad \mathbb{C} \, {\bar
  \psi}_A = \psi_A \quad , \quad A=1,2
\label{gravitinofields}
\end{equation}
In terms of these we define the complex doublet of gravitinos:
\begin{equation}
  \left( \begin{array}{c}
    \psi^\star \\
    \psi \
  \end{array} \right) = \mathcal{C}  \left(\begin{array}{c}
    \psi_1 \\
    \psi_2 \
  \end{array}\right)
\label{compdoub}
\end{equation}
and we introduce the following doublet made by a complex $3$-form current and its complex conjugate:
\begin{equation}
  \mathbb{J}^x_{SU} = \left( \begin{array}{c}
    {\rm i} {\bar \psi}^\star \, \wedge \Gamma_{\underline{a}} \psi \,  \wedge  \, V^{\underline{a}} \\\
     {\rm i} {\bar \psi} \, \wedge \Gamma_{\underline{a}} \psi^\star \,  \wedge  \, V^{\underline{a}} \
  \end{array}\right) \quad \quad (x=\pm)
\label{compcurre}
\end{equation}
By means of an inverse Cayley transformation we get a doublet of real currents:
\begin{equation}
  \mathbb{J}^A_{SL} = \left[ \mathcal{C}^{-1}\right] ^A _{\phantom{A}x} \, \mathbb{J}^x_{SU} =
   \left( \begin{array}{c}
    {\rm i}\left(  {\bar \psi}_1 \, \wedge \Gamma_{\underline{a}} \psi_1 \,  -  \, {\bar \psi}_2 \, \wedge
    \Gamma_{\underline{a}}
    \psi_2
    \right)  \, \wedge  \, V^{\underline{a}} \\
     - 2 {\rm i} {\bar \psi}_1 \, \wedge \Gamma_{\underline{a}} \psi_2 \,  \wedge  \, V^{\underline{a}} \
  \end{array}\right) \equiv d^{A\vert BC} \, {\rm i} \, {\bar \psi}_B
  \, \Gamma_{\underline{a}} \, \psi_C \, \wedge \, V^{\underline{a}}
\label{realdubcur}
\end{equation}
The formula (\ref{realdubcur}) is understood as follows. Recall that
the fermions transform only with respect to the isotropy subgroup
$H=\mathrm{U(1)}\sim \mathrm{O(2)}$ of the scalar coset (are neutral under $G$) and that all irreducible
representations of $\mathrm{O(2)}$ are $2$-dimensional. The
coefficients $d^{A\vert BC}$ defined by equation (\ref{realdubcur})
are the Clebsch Gordon coefficients that extract the doublet of
helicity $s=2$ from the tensor product of two representations of
helicity $s=1$.
Relying on these notations  we can write the type IIB curvature
definitions in two equivalent bases related by a Cayley transformation:
\begin{enumerate}
  \item the complex
$\mathrm{SU(1,1)}$ basis originally used by Castellani and Pesando \cite{igorleo}
  \item the real $\mathrm{SL(2,\mathbb{R})}$, introduced here and
  best suited for comparison with string theory massless modes.
\end{enumerate}
\vskip .4cm
\leftline{\underline{\sl The curvatures of the free differential algebra in the complex
basis}\footnote{Comparing with the original paper by Castellani and
Pesando, note that we have changed the normalization:
$A_{\alpha} \rightarrow \sqrt{2} A_{\alpha}$ and
$B_{\underline{\lambda\mu\nu\rho}} = 6 C_{\underline{\lambda\mu\nu\rho}}$ so that eventually
the $4$--form $C_{[4]}$ will be identified with that used in
Polchinski's book \cite{polchinski}}}
\begin{eqnarray}
   R^{\underline{a}}&=&{\cal D} V^{\underline{a}}-i{\bar\psi}\wedge \Gamma^{\underline{a}}\psi \label{tors1}\\
   R^{\underline{ab}}&=&d \omega^{\underline{ab}}-\omega^{\underline{ac}} \wedge\omega^{\underline{db}} \,
   \eta_{\underline{cd}}\cr
   \rho&=&{\cal D} \psi \equiv d\psi- \ft 1 4  \omega^{\underline{ab}} \wedge \Gamma_{ab}\psi-
          \ft 1 2 {\rm i} Q \wedge \psi \label{lore1}\\
   \mathcal{H}^{\alpha}_{[3]} &=& \sqrt{2} \, dA^{\alpha}_{[2]}+2i\Lambda^{\alpha}_+ {\bar\psi}
   \wedge \Gamma_{\underline{a}} \psi^* \wedge V^{\underline{a}}
          +2{\rm i}\Lambda^{\alpha}_- {{\bar\psi}}^* \wedge
          \Gamma_{\underline{a}} \psi \wedge V^{\underline{a}}\label{2form1}\\
   \mathcal{F}_{[5]} &=& dC_{[4]}+\ft {1} {16} \, {\rm i} \,
   \epsilon_{\alpha\beta} \, \sqrt{2} \, A^{\alpha}_{[2]}  \wedge \mathcal{H}^\beta_{[3]} \,+\, \ft 1 6 \,
   {\bar\psi} \wedge
   \Gamma_{\underline{abc}}\psi \wedge V^{\underline{a}}  \wedge V^{\underline{b}} \wedge
   V^{\underline{c}}
    \nonumber\\
&& + \, \ft 1 8 \, \epsilon_{\alpha\beta} \, \sqrt{2} \, A^{\alpha}_{[2]}
\wedge\left( \Lambda^\beta_+ {\bar\psi} \Gamma_{\underline{a}} \psi^\star +
 \Lambda^\beta_- {\bar\psi}^\star \Gamma_{\underline{a}}
\psi\right) \,
 \wedge V^{\underline{a}} \label{4form1}\\
   {\cal D}\lambda &=& d\lambda-{1\over 4} \omega^{\underline{ab}} \Gamma_{\underline{ab}}\lambda
    -{\rm i}  \ft 3 2 \, Q\lambda
   \label{dilatin1}\\
  {\cal D}\Lambda^{\alpha}_\pm&=&d\Lambda^{\alpha}_{\pm} \mp
   {\rm i}  \, Q \, \Lambda^{\alpha}_{\pm}.\label{su11mc}
\end{eqnarray}
alternatively using the real $\mathrm{SL(2,\mathbb{R})}$ basis we can write:
\vskip .6cm
\leftline{\underline{\sl The curvatures of the free differential algebra in the
real basis}}
\begin{eqnarray}
   R^{\underline{a}}&=&{\cal D} V^{\underline{a}}-i{\bar\psi}_A\wedge \Gamma^{\underline{a}}\psi_A \label{tors2}\\
   R^{\underline{ab}}&=&d \omega^{\underline{ab}}-\omega^{\underline{ac}} \wedge\omega^{\underline{db}} \, \eta_{\underline{cd}}\cr
   \rho_A&=&{\cal D} \psi_A \equiv d\psi_A- \ft 1 4  \omega^{\underline{ab}} \wedge
   \Gamma_{\underline{ab}}\psi_A+
          \ft 1 2  Q \wedge \epsilon_{AB} \psi_B \label{lore2}\\
   \mathbf{H}^{\Lambda}_{[3]}&=&d\mathbf{A}^{\Lambda}_{[3]}+{\rm i}
   \mathbb{L}^{\Lambda}_A \, d^{A\vert BC}
    {\bar\psi}_B \wedge \Gamma_{\underline{a}} \psi_C  \wedge V^{\underline{a}} \label{2form2}\\
   \mathcal{F}_{[5]}&=&dC_{[4]}- \ft {1} {16}
   \epsilon_{\Lambda\Sigma} \mathbf{A}^{\Lambda}_{[3]}  \wedge \mathbf{H}^\Sigma_{[3]} \,+\, {\rm i} \,
   \ft 1 6 \, {\bar\psi}_A \wedge
   \Gamma_{\underline{abc}}\psi_B  \epsilon^{AB} \, V^{\underline{a}} \wedge V^{\underline{b}} \wedge V^{\underline{c}} +
    \nonumber\\
&& + {\rm i} \ft 1 4 \, \epsilon_{\Lambda\Sigma} \mathbf{A}^{\Lambda}_{[2]}
\,  \mathbb{L}^\Sigma_A \, d^{A\vert BC} \,
 {\bar\psi}_B \wedge  \Gamma_{\underline{a}} \psi_C \wedge V^{\underline{a}} \label{4form2}\\
   {\cal D}\lambda &=& d\lambda-{1\over 4} \omega^{\underline{ab}} \Gamma_{\underline{ab}}\lambda- \ft 3 2 {\rm i} Q\lambda\label{dilatin2}\\
  {\cal D} \mathbb{L}^{\Lambda}_{\pm}&=&d\mathbb{L}^{\Lambda}_{A} +
   \epsilon_{AB}Q\mathbb{L}^{\Lambda}_{B}.\label{sl2rmc}
\end{eqnarray}
In the above formulae, eq.s (\ref{su11mc}) and (\ref{sl2rmc}) define the covariant derivative of
the coset representative of the scalar coset in the $\mathrm{SU(1,1)}$ and $\mathrm{SL(2,\mathbb{R})}$ basis
respectively. They follow from the Maurer Cartan equations of $G/H$.
\par
Next, using the results of Castellani and Pesando \cite{igorleo}, we can write the
rheonomic parametrizations of the curvatures
(\ref{tors1}-\ref{su11mc}) (alternatively (\ref{tors2}-\ref{sl2rmc}))
that determine the supersymmetry transformation rules of all the
fields. Prior to that, in order to make contact with  superstring massless modes as normalized in
Polchinski's book, it is convenient to introduce the following
identifications:
\begin{equation}
\begin{array}{rclcrcl}
  \mathbf{A}^1_{[2]} & =& 2 \, \sqrt{2} B_{[2]} & ; & \mathbf{A}^2_{[2]} & =& 2 \, \sqrt{2}
  C_{[2]}\
  \end{array}
\label{identif}
\end{equation}
where $B_{[2]}$ is the $2$--form gauge field of the Neveu-Schwarz
sector that couples to ordinary fundamental strings, while $C_{[2]}$
is the $2$--form of the Ramond-Ramond sector that couples to
$D1$--branes.
For simplicity we write the rheonomic parametrizations only in the complex basis and we disregard the bilinear
fermionic terms calculated by Castellani and Pesando. We have:
\begin{eqnarray}
 R^{\underline{a}}&=&0\nonumber\\&{}\label{torpar}\\
 \rho&=&\rho_{\underline{ab}} V^{\underline{a}} \wedge V^{\underline{b}}
      + {5\over 16}{\rm i}\Gamma^{\underline{a_1-a_4}} \psi V^{\underline{a_5}} \left(
          F_{\underline{a_1-a_5}}
          +{1\over5!}\epsilon_{\underline{a_1-a_{10}}}F_{\underline{a_6-a_{10}}}
        \right)\nonumber\\
     & &+{1\over 32}\left(
         -\Gamma^{\underline{a_1-a_4}}\psi^*V_{\underline{a_1}}
         +9\Gamma^{\underline{a_2a_3}}\psi^*V^{\underline{a_4}}
       \right)\Lambda^{\alpha}_+\mathcal{H}^{\beta}_{\underline{a_2-a_4}} \epsilon_{\alpha\beta}\nonumber\\
     & &+\mbox{fermion bilinears}\label{rhopar}\\
&{}&\nonumber\\
 \mathcal{H}^{\alpha}_{[3]}&=&
     \mathcal{H}^{\alpha}_{\underline{abc}} V^{\underline{a}} \wedge V^{\underline{b}} \wedge
     V^{\underline{c}}
    +\Lambda^{\alpha}_+ {\bar \psi}^* \Gamma_{\underline{ab}}\lambda^* V^{\underline{a}} \wedge
    V^{\underline{b}}
    +\Lambda^{\alpha}_- {\bar \psi} \Gamma_{\underline{ab}} \lambda V^{\underline{a}} \wedge V^{\underline{b}}
    \label{Hpar}\\
&{}&\nonumber\\
\mathcal{F}_{[5]}&=&F_{\underline{a_1-a_5}}V^{\underline{a_1}}
\wedge \dots \wedge V^{\underline{a_5}}\label{f5par}\\&{}\nonumber\\
{\cal D} \lambda&=&
   {\cal D}_{\underline{a}}\lambda V^{\underline{a}}
   +{\rm i} P_{\underline{a}} \Gamma^{\underline{a}} \psi^*
   -{1 \over 8}{\rm i}\Gamma^{\underline{a_1-a_3}} \psi
      \epsilon_{\alpha\beta}\Lambda^{\alpha}_+\mathcal{H}^{\beta}_{\underline{a_1-a_3}}
      \label{dilatinpar}\\&{}&\nonumber\\
{\cal D}\Lambda^{\alpha}_+&=&
    \Lambda^{\alpha}_- P_{\underline{a}}V^{\underline{a}}
    + \Lambda^{\alpha}_-{\bar \psi}^* \lambda
\label{GsuHpar1}\\&{}&\nonumber\\
{\cal D}\Lambda^{\alpha}_-&=&
   \Lambda^{\alpha}_+P_{\underline{a}}^*V^{\underline{a}}
   + \Lambda^{\alpha}_+{\bar \psi}\lambda^*
\label{GsuHpar2}\\
R^{\underline{ab}}&=&R^{\underline{ab}}_{~~\underline{cd}} V^{\underline{c}} \wedge V^{\underline{d}} \,
+ \, \mbox{fermionic
terms}\label{lorenpar}
\end{eqnarray}
\subsection{The bosonic field equations and the standard form of the bosonic
action}
Following Castellani and Pesando we write next the general form of
the bosonic field equations and using the identifications of eq.s
(\ref{identif}), (\ref{Lfatcoset}), (\ref{PQvalue}) we reduce them to
those following from a {\sl standard supergravity  action for $p$--branes}.
As discussed in the literature \cite{stellebrane,parilez,bholelec},
the standard form of a supergravity action truncated to the graviton, the dilaton and
the $n_i=p_i+2$ field strengths that can couple to the world--volume
actions of $p_i$--branes is as follows:
\begin{eqnarray}
  \mathcal{A}_{standard} & = &\int \, d^Dx \, \mbox{det}V \, \Biggl [
  -2 \, R\left[ \omega(V)\right] - \ft 1 2 \, \partial ^{\underline{\mu} } \varphi \,
  \partial _{\underline{\mu}}  \varphi \Biggr ] \nonumber\\
 &&  - \int \, \sum_i \,\frac{1 }{2 } \, \exp\left [
  - a_i \,\varphi\right]  \,  F_{[n_i]} \, \wedge \, \star \, F_{[n_i]}  \, + \,
  \mbox{Chern Simons couplings}
\label{standaction}
\end{eqnarray}
where $ R= R^{\underline{ab}}_{\underline{ab}} $ is the scalar curvature in the geometric
normalizations always adopted in the rheonomic framework \cite{castdauriafre} \footnote{Note
that our $R$ is equal to $- \ft 1 2 R^{old}$, $R^{old}$ being the
normalization of the scalar curvature usually adopted in General
Relativity textbooks. The difference arises because in the traditional literature the
Riemann tensor is not defined as the components of the curvature
$2$--form $R^{ab}$ rather as $-2$ times such components}, and $a_i$
are characteristic exponents dictated by the structure of
supergravity and playing an essential role in dictating the
properties of $p$--brane solutions. Furthermore in eq.
(\ref{standaction}) we have defined:
\begin{eqnarray}
  \vert \, F_{[n]} \vert ^2 & \equiv &  g^{\underline{\mu_1\nu_1}} \, \dots
  g^{\underline{\mu_{n}\nu_n}} \, F_{\underline{\mu _1 \dots \mu _n}} \, F_{\underline{\nu _1 \dots \nu
  _n}} \label{modFn}\\
F_{[n]} & = & F_{\underline{\mu _1 \dots \mu _n}} \, dx^{\underline{\mu _1}} \, \wedge \,
\dots \wedge dx^{\underline{\mu _n}}
\label{normalform}
\end{eqnarray}
and we have not made explicit the Chern Simons couplings between field
strengths that are on the other hand essential in the derivation of the exact
field equations.
\par
Introducing the definition of the dressed $3$--form field strengths:
\begin{equation}
  \begin{array}{rclcrcl}
  \widehat{\mathcal{H}}_{\pm \vert \underline{a_1 a_2 a_3}} & = & \epsilon_{\alpha\beta}
  \Lambda^\alpha_\pm \, \mathcal{H}_{\underline{a_1 a_2 a_3}} & ; &
  \widehat{\mathbf{H}}_{A\vert \underline{a_1 a_2 a_3}} & = &
  \epsilon_{\Lambda\Sigma} \, \mathbb{L}^\Lambda_{\phantom{\Lambda}
  A} \, \mathbf{H}^\Sigma_{\underline{a_1 a_2 a_3}} \
  \end{array}
\label{dressedfielstr}
\end{equation}
it was shown by Castellani and Pesando \cite{igorleo} that the exact bosonic field
equations implied by the closure of the supersymmetry algebra have the following form:
\begin{eqnarray}
R^{\underline{pr}}_{~~\underline{qr}}-
\ft 1 2 \delta^{\underline{p}}_{\underline{q}} R^{\underline{ab}}_{~~\underline{ab}}&=&
-{75}\left(
     F_{\underline{qa_1-a_4}}F^{\underline{pa_1-a_4}}-{1\over 10}
     \delta^{\underline{p}}_{\underline{q}}
   F_{\underline{a_1-a_5}}F^{\underline{a_1-a_5}}
   \right)
 \nonumber \\
  & &-{9\over 16} \left(
    \widehat{\mathcal{H}}_{+}^{\underline{pa_1a_2}} \widehat{\mathcal{H}}_{- \vert \underline{qa_1a_2}}+
    \widehat{\mathcal{H}}_{-}^{\underline{pa_1a_2}} \widehat{\mathcal{H}}_{+ \vert \underline{qa_1a_2}}
     -{1\over 3} \delta^{\underline{p}}_{\underline{q}}
     \widehat{\mathcal{H}}_{+}^{\underline{a_1a_2a_3}}\widehat{\mathcal{H}}_{- \vert \underline{a_1a_2a_3}}
   \right)
 \nonumber \\
  & &-\ft 1 2\left(
     P^{\underline{p}}P^*_{\underline{q}}+P_{\underline{q}}P^{*\underline{p}}-
     \delta^{\underline{p}}_{\underline{q}} P^{\underline{a}} P^*_{\underline{a}}
    \right)
\label{einsteineq}  \\
 & &\nonumber\\
 {\cal D}^{\underline{a}} P_{\underline{a}}&=&
   -\ft 3 8 \,  \widehat{\mathcal{H}}_{+}^{\underline{a_1a_2a_3}}\widehat{\mathcal{H}}_{+\vert \underline{a_1a_2a_3}}
\label{scalareq}\\& &\nonumber\\
   {\cal D}^{\underline{b}}\widehat{\mathcal{H}}_{+\vert \underline{a_1a_2 b}}&=&
    - {\rm i} 20 \, F_{\underline{a_1 a_2 b_1 b_2 b_3}} \widehat{\mathcal{H}}_{+}^{\underline{b_1 b_2 b_3}}
    -P^{\underline{b}} \widehat{\mathcal{H}}_{- \vert \underline{a_1a_2 b}}
\label{2formeq}\\& &\\
   {\cal D}^{\underline{b}} F_{\underline{a_1a_2 a_3 a_4 b}}&=&
   {\rm i} \ft {1} {960} \epsilon_{\underline{a_1a_2 a_3 a_4 b_1 \dots b_6}}
      \widehat{\mathcal{H}}_{+}^{\underline{ b_1 b_2 b_3}} \widehat{\mathcal{H}}_{-}^{\underline{b_4 b_5 b_6}}
\label{5formequa}
\end{eqnarray}
At the purely bosonic level (i.e. disregarding all fermionic
contributions), using the solvable parametrization (\ref{Lfatcoset})
of the $\mathrm{SL(2,\mathbb{R})/O(2)}$ coset and inserting the
identifications (\ref{identif}) we obtain the following expression
for the dressed $3$--forms in terms of string massless fields denoted $NS$ or $RR$ according to their origin
in the Neveu Schwarz or Ramond Ramond sector:
\begin{eqnarray}
\widehat{\mathcal{H}}_\pm & = & \pm 2 \,e^{-\varphi/2} F^{NS}_{[3]} + {\rm i} 2
\,e^{\varphi/2} \,F^{RR}_{[3]}
\,  \nonumber\\
P & =& \ft 1 2 \, d\varphi -{\rm i} \ft 12 \, e^\varphi \, F_{[1]}^{RR} \nonumber\\
F^{NS}_{[3]} & = & dB_{[2]} \nonumber\\
F^{RR}_{[1]} & = & dC_{[0]} \nonumber\\
F^{RR}_{[3]}& = & \left( dC_{[2]} -  \, C_{[0]} \,
dB_{[2]}\right)\nonumber\\
F^{RR}_{[5]}& = &  \mathcal{F}_{[5]} = dC_{[4]}-
\ft 12 \left( B_{[2]} \wedge d C_{[2]} -  C_{[2]} \wedge d B_{[2]}\right)
\nonumber\\
\label{defihpm}
\end{eqnarray}
Using the convention (\ref{hodgedual}) for the Hodge dual of $\ell$--forms in
space--time dimensions $D$,
the field equations (\ref{scalareq}-\ref{5formequa}) can be written
in more compact form. Let us begin with the scalar equation
(\ref{scalareq}), it becomes:
\begin{equation}
  d(\star P)- 2\, {\rm i} \, Q \wedge \star P +
\ft 1 {16}
   \widehat{\mathcal{H}}_+ \, \wedge \, \star
  \widehat{\mathcal{H}}_+=0
\label{dstarP}
\end{equation}
and separating its real from imaginary part we obtain the two
equations:
\begin{eqnarray}
d \star d \varphi - e^{2\varphi} \, F^{RR}_{[1]} \wedge F^{RR}_{[1]} & = & -\ft 1 2 \,
\left( e^{-\varphi} F^{NS}_{[3]} \wedge \star  F^{NS}_{[3]}-
  e^{\varphi} F^{RR}_{[3]} \wedge \star  F^{RR}_{[3]}\right) \label{NSscala}\\
d\left( e^{2\varphi} *F^{RR}_{[1]}\right)  & = & - e^{\varphi} \, F^{NS}_{[3]} \wedge \star  F^{RR}_{[3]}
\label{RRscala}
\end{eqnarray}
Considering next the $3$--form eq. (\ref{2formeq}) it can be
rewritten as:
\begin{equation}
  d \star \widehat{\mathcal{H}}_+ - {\rm i} \, Q \wedge \, * \widehat{\mathcal{H}}_+=
   {\rm i} \, \mathcal{F}_{[5]}\,
  \wedge \, \widehat{\mathcal{H}}_+  - P \wedge \star
  \widehat{\mathcal{H}}_-
\label{hodge2formeq}
\end{equation}
Separating the real and imaginary parts of eq.(\ref{hodge2formeq}) we obtain:
\begin{eqnarray}
d\left( e^{-\varphi} \, \star F_{[3]}^{NS}\right) + e^\varphi \, F^{RR}_{[1]} \wedge \star F^{RR}_{[3]}
  & = &  - F_{[3]}^{RR} \wedge F^{RR}_{[5]}
\nonumber\\
d\left( e^\varphi \star F_{[3]}^{RR
} \right) & = & -F_{[5]}^{RR} \, \wedge F_{[3]}^{NS}
\label{3formequazie}
\end{eqnarray}
Finally the equation for the Ramond-Ramond $5$--form, namely equation \ref{5formequa} is rewritten as follows:
\begin{equation}
  d\star F^{RR}_{[5]} = {\rm i} \, \ft 1 {8} \, \widehat{\mathcal{H}}_+ \wedge \widehat{\mathcal{H}}_- = -
  F^{NS}_{[3]} \, \wedge \, F^{RR}_{[3]}
\label{f5equazia}
\end{equation}
\section{The $\kappa$-supersymmetry projector}
\label{projectap}
Let us begin with property $\mbox{[a]}$ and consider the ansatz in
eq.(\ref{def gamma}). By direct calculation we find:
\begin{eqnarray}
\omega_{[4]}^{2} \, & = & \, \alpha_{4}^{2} (4!)^{2} \nonumber \\
\omega_{[2]}^{2} \, & = & \, \frac{(\alpha_{2})^{2} \, \omega_{[0]} \,
\omega_{[4]} }{3! \, \alpha_{0} \, \alpha_{4} } \, + \, 8 (\alpha_{2})^{2} Tr(\hat{F}^{2})
\label{def omega quad}
\end{eqnarray}
so that we get:
\begin{equation}
\Gamma^{2} = \frac{1}{N^{2}} \, \left[ (4! \alpha_4)^2 \, + \, \omega_{[0]} \omega_{[4]} \left(\frac{(\alpha_2)^{2}}{3!\alpha_0 \alpha_4} + 2 \right) \, + \, 8(\alpha_2)^2 \, Tr(\hat{F}^2) \right]
\label{bamma quadro}
\end{equation}
so  we obtain $\Gamma^{2} = 1 \! \!  1 $
if the normalization factor $N$ is chosen as in eq. (\ref{def omega})
and if the coefficients are chosen as in eq. (\ref{def alfa}). This
conclusion is easily reached using the identity (\ref{determinante
eta}) of the main text.
\par
Let us now turn to property $\mbox{[b]}$, namely to the condition
\begin{equation}
 \Gamma \, A_{k} \, = \,  A_{k}
\label{k supersi}
\end{equation}
To implement it we need to calculate some $\gamma$ matrix  products:
\begin{eqnarray}
\omega_{[4]} \, \gamma_{k} & = & \frac{1}{6} \, \tilde{\gamma} \,_{k} \nonumber \\
\omega_{[4]} \, \tilde{\gamma} \,_{k} & = & 6 \, \gamma_{k} \nonumber \\
\omega_{[4]} \, \Pi_{k} & = & - \, \frac{1}{2} \, \tilde{\hat{F}}^{ij}\gamma_{ijk} \equiv - \frac{1}{2}
\, \tilde{\Delta}_{k} \nonumber \\
\omega_{[4]} \tilde{\Pi} \, _{k} & = & - \, \frac{1}{2} \hat{F}^{ij}\gamma_{ijk} \equiv - \frac{1}{2} \, \Delta_{k}
\label{prod not omega 4}
\end{eqnarray}
\begin{eqnarray}
\omega_{[2]} \, \gamma_{k} & = & \frac{i}{2} \tilde{\Delta}_{k} +
i \, \tilde{\Pi} \, _{k} \nonumber \\
\omega_{[2]} \, \tilde{\gamma} \,_{k} & = & - 3 i \, \Delta_{k} - 6 i \, \Pi_{k} \nonumber \\
\omega_{[2]} \, \Pi_{k} & = & - \frac{i}{6} \, (\hat{F}^{2})_{kl} \, \tilde{\gamma} \,^{l} - i \,
\omega_{[0]} \, \gamma_{k} \nonumber \\
\omega_{[2]} \tilde{\Pi} \, _{k} & = & i \,(\tilde{\hat{F}} \, ^{2})_{kl} \, \gamma^{l}
+ \frac{i}{6} \, \omega_{[0]} \, \tilde{\gamma} \,_{k}
\label{prod not omega 2}
\end{eqnarray}
Now we impose equation (\ref{k supersi}) and we obtain the
following equations.\\
$\bullet$ The contributions from $\Delta_{k}$ and $\tilde{\Delta}_{k}$ are :
\begin{eqnarray}
\Delta^{m} \, h_{mk} \, (\frac{i}{2} \, f_{4} + 3 \, f_{2}) \otimes \sigma_{1}
& = & 0 \nonumber \\
\tilde{\Delta}_{k} \, ( - \frac{i}{2} \, f_{3} + \frac{i}{2} \, f_{1} )
\otimes \sigma_{2} & = & 0
\label{delta contr}
\end{eqnarray}
$\bullet$ For the contributions with $\gamma_k$ we have two equations, one proportional to $\sigma_{3}$
and one proportional to $ 1 \! \! 1$, namely :
\begin{equation}
\gamma_{k} \, \omega_{[0]} \, ( f_{1} - f_{3} ) \otimes \sigma_{3} = 0
\label{gamma1 contr}
\end{equation}
and
\begin{eqnarray}
\frac{1}{N} \, \left(6 \, f_{2} 1 \! \! 1_{4 \times 4} + i \, f_{4}
\tilde{\hat{F}}^{2} \right) \, h \otimes 1 \! \! 1_{2 \times 2} = f_1 \,
1 \! \! 1_{4 \times 4} \otimes 1 \! \! 1_{2 \times 2}
\label{gamma2 contr}
\end{eqnarray}
For:
\begin{eqnarray}
6 \, f_{2} = f_1 \nonumber \\
f_1 = -i \, f_4
\label{vinc gamma2}
\end{eqnarray}
and using the property (\ref{prop 2 3}) we obtain:
\begin{eqnarray}
N^{-1} \, \left(1 \! \! 1 - \tilde{\hat{F}}^{2} \right) \, h & = & 1 \! \! 1 \nonumber \\
N^{-1} \, \left(1 \! \! 1 - \tilde{\hat{F}}^{2} \right) \, N^{-1} \left(1 \! \! 1 - \hat{F}^{2} \right)
& = & 1 \! \! 1 \nonumber \\
\left( 1 \! \! 1 - \hat{F}^{2} - \tilde{\hat{F}}^{2} + \tilde{\hat{F}}^{2} \hat{F}^{2}\right)
& = & N^{2} \nonumber\\
\left[ 1 \! \! 1 - \frac{1}{2} \, Tr(\hat{F}^{2}) + \left(\frac{1}{8} F_{ij} \, F_{kl} \,
\epsilon^{ijkl} \right) \, 1 \! \! 1 \right] & = & N^{2}
\label{gamma3 contr}
\end{eqnarray}
$\bullet$ For the contributions with $\tilde{\gamma}_k$ we get the following equations:
\begin{eqnarray}
\tilde{\gamma}_{m} \, h^{m} \, \,_{k} \, \omega_{[0]} \left( f_{2} + \frac{i}{6} f_4 \right)
\otimes 1 \! \! 1_{2 \times 2} & = & 0 \nonumber \\
N^{-1} \, \tilde{\gamma}_{m} \, \left[ \frac{1}{6} \, f_1 \, \delta^{m} \, _{k}
-\frac{1}{6} \, f_3 \, (\hat{F}^{2})^{m}\,_{k}  \right] \otimes \sigma_3 & = & f_2 \,
\tilde{\gamma}_{m} \, h^{m} \, _{k} \otimes \sigma_3
\label{gamma tilde contr}
\end{eqnarray}
Then if :
\begin{eqnarray}
f_{2} & = & - \frac{i}{6} \, f_{4} \nonumber \\
f_{1} & = & f_3 \nonumber \\
f_1 & = & 6 f_2
\label{const susy2}
\end{eqnarray}
we obtain  that the second of equations (\ref{gamma tilde contr}) as a matrix equation becomes:
\begin{eqnarray}
N^{-1} \, [1 \! \! 1 - (\hat{F}^{2})] = h
\label{gamma tilde 2 contr}
\end{eqnarray}
and just coincides with the solution (\ref{h no eta}) for the auxiliary field $h$ in terms of the physical ones.\\
$\bullet$ Now we consider $\Pi$ and $\tilde{\Pi}$.\\
The equation proportional to $\sigma_1$ is :
\begin{eqnarray}
\alpha \, \omega_{[0]} \tilde{\Pi}^{m} \, h_{mk} \, + \, \beta \, \Pi^{m} \, h_{mk} \, & = & \, N \,
\gamma \, \Pi^{k} \nonumber \\
\alpha \, \omega_{[0]} \tilde{\hat{F}}^{lm}  \, h_{mk} \, \gamma_{l} + \beta \hat{F}^{lm} \, h_{mk} \,
\gamma_{l}  & = & \gamma \,N \, \hat{F}^{l} \,_{k} \, \gamma_{l}
\label{pi pitilde 1}
\end{eqnarray}
in matrix form we have :
\begin{eqnarray}
\alpha \, \omega_{[0]} \, (\tilde{\hat{F}} \, h) + \beta \, (\hat{F} \, h) & = & \gamma \, N \, \hat{F} \nonumber \\
\alpha \, \omega_{[0]} \, \tilde{\hat{F}} \, [1 \! \! 1 - (\hat{F}^{2})] \,N^{-1} + \beta \, \hat{F}^{2} \, [1 \! \! 1
 - (\hat{F}^{2})] \,N^{-1} & = & \gamma \, N \, \hat{F} \nonumber \\
\alpha \, \omega_{[0]} \, \tilde{\hat{F}} \, [1 \! \! 1 - (\hat{F}^{2})] \, + \beta \, \hat{F}^{2} \, [1 \! \! 1
 - (\hat{F}^{2})]  & = & \gamma \, N^{2} \, \hat{F} \nonumber \\
\alpha \, \omega_{[0]} \, \tilde{\hat{F}} \, [1 \! \! 1 - (\hat{F}^{2})] \, + \beta \, \hat{F}^{2} \, [1 \! \! 1
- (\hat{F}^{2})]  & = & \gamma \, [1 -
\frac{1}{2} \, Tr(\hat{F}^{2}) + \omega_{[0]}^{2}] \, \hat{F} \nonumber \\
\alpha \, \omega_{[0]} \, \tilde{\hat{F}} - \alpha \, \omega_{[0]} \,
(\tilde{\hat{F}}\hat{F})\hat{F} + \beta \, \hat{F} - \beta \, \hat{F}^{2}\hat{F} & = & \gamma \, \hat{F}
- \frac{\gamma}{2} \, Tr(\hat{F}^{2})\hat{F} + \gamma \, \omega_{[0]}^{2} \, \hat{F}
\label{pi pitilde 1 matr}
\end{eqnarray}
if :
\begin{eqnarray}
\beta = \gamma
\label{identita 10}
\end{eqnarray}
and using (\ref{prop 2 3}), than (\ref{pi pitilde 1 matr}) become :
\begin{eqnarray}
\alpha \, \omega_{[0]} \, \tilde{\hat{F}} + \alpha \, \omega_{[0]}^{2} \, \hat{F} - \beta \, \hat{F}^{2}\hat{F} =
 - \frac{\gamma}{2} \, Tr(\hat{F}^{2})\hat{F} + \gamma \, \omega_{[0]}^{2} \, \hat{F}
\label{pi pitilde 2  matr}
\end{eqnarray}
if :
\begin{eqnarray}
\alpha = \gamma
\label{identita 12}
\end{eqnarray}
\begin{eqnarray}
\alpha \, \omega_{[0]} \, \tilde{\hat{F}} - \beta \, \hat{F}^{2}\hat{F} & = &
- \frac{\gamma}{2} \, Tr(\hat{F}^{2})\hat{F} \nonumber \\
\alpha \, \omega_{[0]} \, \tilde{\hat{F}} - \alpha \, \hat{F}^{2}\hat{F} & = &
- \alpha \, (\hat{F}^{2} + \tilde{\hat{F}}^{2})\hat{F} \nonumber \\
\alpha \, \omega_{[0]} \, \tilde{\hat{F}} & = & - \alpha \tilde{\hat{F}}^{2}\hat{F}
\label{pi pitilde 3  matr}
\end{eqnarray}
and it is correct by (\ref{prop 2 3}).\\
The equation proportional to $\sigma_{2}$ is :
\begin{eqnarray}
\mu \, \omega_{[0]} \, \Pi_{k} + \nu \, \tilde{\Pi}_{k} & = & N \, \rho \, \tilde{\Pi}^{m} \, h_{mk} \nonumber \\
\mu \, \omega_{[0]} \, \hat{F}_{lk} \, \gamma^{l} + \nu \, \tilde{\hat{F}}_{lk} \, \gamma^{l} & = & N \,
\rho \, \tilde{\hat{F}}_{l} \,^{m} \, h_{mk} \, \gamma^{l} \nonumber \\
\mu \, \omega_{[0]} \, \hat{F}_{lk} + \nu \, \tilde{\hat{F}}_{lk} & = &
 N \, \rho \, \tilde{\hat{F}}_{lm} \, (\delta^{m} \,_{k} - [\hat{F}^{2})^{m} \,_{k}] \, N^{-1}
\label{pi pitilde 4  matr}
\end{eqnarray}
for :
\begin{eqnarray}
\nu = \rho \nonumber \\
\mu = \rho
\label{vinc 7}
\end{eqnarray}
we obtain the first of the relations (\ref{prop 2 3}).\\
Where :
\begin{eqnarray}
\alpha = -i \, f_{4}  \,\,\,\,\,\,  \beta = 6 i f_{2}  \,\,\,\,\, \,   \gamma = f_{3} \nonumber \\
\mu = i \, f_{3}  \,\,\,\,\,\,  \nu = i \, f_{1}  \,\,\,\,\,\,  \rho = f_{4}
\label{mu nu ..}
\end{eqnarray}
Using the fact that $a_{5}= \frac{3}{4}i$ and (\ref{gli f}) we have that
(\ref{delta contr}), (\ref{gamma1 contr}), (\ref{vinc gamma2}),
 (\ref{const susy2}), and (\ref{identita 10}),
(\ref{identita 12}), (\ref{vinc 7}) are automatically satisfied.
This concludes the proof of property $\mbox{[b]}$ and hence of $\kappa$
supersymmetry
\section{Notations and Conventions}
\label{notazie}
General adopted notations for first order actions are the
following ones:
\begin{eqnarray}
d & = & \mbox{ dimension of the world-volume $ \mathcal{W}_d$} \nonumber\\
D & = & \mbox{dimension of the bulk space--time $\mathcal{M}_D $ }\nonumber\\
V^{\underline{a}}&=& \mbox{vielbein $1$--form of bulk
space--time}\nonumber\\
\Pi^{\underline{a}}_{i} & = & D\times d \,\, \mbox{matrix. $0$--form auxiliary
field}\nonumber\\
h^{ij} &=& d\times d \,\,\mbox{symmetric matrix. $0$--form auxiliary
field}\nonumber\\
e^{\ell} &=& \mbox{vielbein $1$--form of the
world-volume}\nonumber\\
\eta_{\underline{ab}} & = & \mbox{diag}\{ +,\underbrace{-,\dots,-} _{D-1 \, times}
\} = \mbox{flat metric on the bulk}\nonumber\\
\eta^{\underline{ij}} & = & \mbox{diag} \{ +,\underbrace{-,\dots,-} _{d-1 \, times}
\} = \mbox{flat metric on the world--volume}\nonumber\\
\label{definizie}
\end{eqnarray}
The supersymmetric formulation of type IIB supergravity we rely on is
that of Castellani and Pesando \cite{igorleo} that uses the rheonomy
approach \cite{castdauriafre}. Hence, as it is customary in all the
rheonomy constructions, the adopted signature of space--time is the
\textbf{mostly minus signature}:
\begin{equation}
  \eta_{\underline{ab}}= \mbox{diag} \left\{ + ,\underbrace{
  -,\dots,-}_{\mbox{9 times}} \right \}
\label{mostmin}
\end{equation}
The index conventions are the following ones:
\begin{eqnarray}
\underline{a,b,c,\dots} & = & 0,1,2,\dots,9 \quad \mbox{Lorentz flat indices in $D=10$} \nonumber\\
i,j,k,\dots & = & 0,\dots ,d  \quad \mbox{Lorentz flat indices on the world-volume}\\
\alpha, \beta, \dots & = &  1,2  \quad \mbox{$\mathrm{SU(1,1)}$ doublet
indices}\nonumber\\
A,B,C,\dots & = & 1,2 \quad \mbox{$\mathrm{O(2)}$ indices for the scalar
coset}\nonumber\\
\Lambda,\Sigma,\Gamma,\dots &=& 1,2 \quad \mbox{$\mathrm{SL(2,R)}$ doublet
indices}
\label{indeconv}
\end{eqnarray}
For the gamma matrices our conventions are as follows:
\begin{equation}
  \left\{ \Gamma^{\underline{a}} \, , \, \Gamma^{\underline{b}} \right\}  = 2\,  \eta^{\underline{ab}}
\end{equation}
The convention for
constructing the dual of an $\ell$--form $\omega$ in D--dimensions is
the following:
\begin{equation}
  \omega= \omega_{\underline{i_1\dots i_\ell}} \, V^{\underline{i_1}} \wedge \dots \wedge
  V^{\underline{i_\ell}} \quad \Leftrightarrow \quad \star \omega = \frac{1}{(D-\ell)!} \,
  \epsilon_{\underline{a_1 \dots a_{D-\ell} b_1
  \dots b_\ell}} \omega^{\underline{b_1\dots
  b_\ell}} \, V^{\underline{a_1}} \wedge \dots \wedge V^{\underline{a_{D-\ell}}}
\label{hodgedual}
\end{equation}
Note that we also use $\ell$--form components with strength one:
$\omega= \omega_{\underline{i_1\dots i_\ell}} \, V^{\underline{i_1}} \wedge \dots \wedge
  V^{\underline{i_\ell}}$ and not with strength $\ell!$ as it would
  be the case if we were to write $\omega= \frac {1}{\ell !}\omega_{\underline{i_1\dots i_\ell}} \,
  V^{\underline{i_1}} \wedge \dots \wedge V^{\underline{i_\ell}}$
When it is more appropriate to use curved rather than flat indices
then the convention for Hodge duality is summarized by the formula:
\begin{equation}
  \star\left(dx^{\mu_1}\wedge\dots dx^{\mu_n}\right)=\frac{\sqrt{-{\rm det}(g)}}{(10-n)!}G^{\mu_1\nu_1}\dots
G^{\mu_n\nu_n}\epsilon_{\rho_1\dots
 \rho_{10-n}\nu_1\dots \nu_n}\,dx^{\rho_1}\wedge\dots dx^{\rho_{10-n}}
\label{hodgecurvo}
\end{equation}

\end{document}